\documentclass[aps,pra,twocolumn,reprint,superscriptaddress,tightenlines,showpacs,longbibliography]{revtex4-2}
\usepackage{graphicx}
\usepackage{float}
\usepackage{amsmath}
\usepackage{amssymb}
\usepackage{array}
\usepackage{bm}
\usepackage{color}
\usepackage{siunitx}
\usepackage[colorlinks,linkcolor=blue,anchorcolor=blue,citecolor=pink2,urlcolor=blue]{hyperref}
\definecolor{pink2}{RGB}{220, 60, 110}
\usepackage{multirow}
\usepackage{multibib}
\usepackage[shortlabels]{enumitem}
\DeclareMathOperator{\Tr}{\mathrm{Tr}}
\newcommand{\PT}{$\mathcal{PT}$}
\newcommand{\A}{$\mathcal{A}$}
\newcommand{\blue}{\textcolor{blue}}
\newcommand{\NUDTQIS}{Institute for Quantum Science and Technology, College of Science, NUDT, Changsha 410073, Hunan, China}
\newcommand{\NUDTQKL}{Hunan Key Laboratory of Quantum Information Mechanism and Technology, NUDT, Changsha 410073, Hunan, China}
\newcommand{\GXUST}{College of Science, Guangxi University of Science and Technology, Liuzhou 545006, Guangxi, China}

\begin{document}
\title{\PT-symmetric quantum sensing: advantages and restrictions}
\author{Yan-Yi Wang}
\email{wangyy@gxust.edu.cn}
\affiliation{\NUDTQIS}
\affiliation{\GXUST}
\affiliation{\NUDTQKL}
\author{Chun-Wang Wu}
\email{cwwu@nudt.edu.cn}
\affiliation{\NUDTQIS}
\affiliation{\NUDTQKL}
\author{Wei Wu}
\affiliation{\NUDTQIS}
\affiliation{\NUDTQKL}
\author{Ping-Xing Chen}
\affiliation{\NUDTQIS}
\affiliation{\NUDTQKL}

\date{\today}

\begin{abstract}
Quantum sensing utilizing unique quantum properties of non-Hermitian systems to realize ultra-precision measurements has been attracting increasing attention.
However, the debate on whether non-Hermitian systems are superior to Hermitian counterparts in sensing remains an open question.
Here, we investigate the quantum information in \PT-symmetric quantum sensing utilizing two experimental schemes based on the trapped-ion platform.
It turns out that the existence of advantages of non-Hermitian quantum sensing heavily depends on additional information resources carried by the extra degrees of freedom introduced to construct \PT-symmetric quantum sensors.
Moreover, the practical application of non-Hermitian quantum sensing with superior performance is primarily restricted by the additional resource consumption accompanied by the post-selection.
Our study provides theoretical references for the construction of non-Hermitian quantum sensors with superior performance and has potential applications in research fields of quantum precision measurement and quantum information processing.
\end{abstract}
%\pacs{}
\maketitle

\section{\label{sec1} Introduction}
In conventional quantum mechanics, the fundamental assumption of Hermiticity guarantees the reality of the eigenvalues of Hamiltonians and the unitarity of the time evolution of quantum systems.
However, in the last two decades, \textsl{non-Hermitian} (NH) quantum mechanics has received considerable attention.
Bender \emph{et al.} proposed that \PT-symmetric systems possessing non-Hermiticity can maintain real and positive energy spectra \cite{PhysRevLett.80.5243,RepProgPhys.70.947}, and redefined the \PT~inner product \cite{PhysRevLett.89.270401}.
\PT-symmetric systems also can be reinterpreted as a NH subsystem embedded into a larger Hermitian system \cite{PhysRevLett.101.230404,PhysRevA.78.042115,PhysRevA.91.052113,PhysRevLett.119.190401}.
Mostafazadeh further introduced the pseudo-Hermiticity and explored a more general sense of \PT-symmetry \cite{JMathPhys.43.205,JMathPhys.43.2814}.
The development of NH physics has been greatly propelled by the proposal of \PT-symmetry, and its applications have been widely extended across various research branches of quantum science and technology \cite{NHQM2011Moiseyev,NatPhys.14.11,AdvanPhys.69.249}.
Particularly, the \textsl{exceptional point} (EP) of \PT-symmetric Hamiltonian involves a switch in the eigenvalues from real to complex accompanying the degeneracy of the eigenstates \cite{Sci.363.7709,PhysRevA.100.062131,RevModPhys.93.015005}, in consequence, \PT-symmetry is spontaneously broken.
Hence, the EP-induced degeneracy, caused by the sensitivity divergence of the eigenenergy spectral gap \cite{Nat.548.192}, becomes one of the most striking characteristics of NH systems \cite{Sci.346.328,Nat.537.80,NatCommun.9.2674}, and the observation of \PT-symmetry spontaneous breaking has been realized in various experimental works \cite{PhysRevLett.103.093902,Sci.364.878,NatCommun.10.855,NatPhys.15.1232}.
Recent studies, both in theory \cite{PhysRevLett.117.110802,NatCommun.11.5382,PhysRevLett.124.020501,PhysRevLett.128.173602} and experiment \cite{Nat.548.192,Nat.548.187,Nat.576.65}, have reported that the EP can enhance both quantum and (quasi-) classical sensing \cite{PhotonRes.8.1457}.\\

\textsl{Quantum sensing} (QS) combined with precision measurement and quantum physics exploits unique quantum features to achieve ultra-precision measurement \cite{QMC2010Wiseman,RevModPhys.89.035002}.
Within the framework of quantum theory, developing precise measurement methods for surpassing the limitations of classical techniques by leveraging quantum resources is currently one of the main goals in this research field.
In the last dozen years, QS has made rapid development in various systems \cite{NatPhot.5.222}, including single photon \cite{NatPhot.12.724,PhysRevLett.125.240506}, cold atoms \cite{RevModPhys.82.2313,Nat.604.457}, trapped ions \cite{Nat.603.604,PhysRevLett.126.083604}, superconducting qubits and solid-state spins \cite{PhysRevLett.125.117701,RevModPhys.92.015004,npjQInfo.9.56}.
Our study is based on the trapped-ion platform which has certain advantages such as long coherence times, multiple vibrational modes, and robustness against environmental disturbances, making it suitable for QS \cite{Nat.603.604}.
Recently, a debate has arisen about whether NH physics is superior in sensing.
Except for the studies of \textsl{EP-enhanced quantum sensing} (EPQS)
\cite{Nat.548.192,PhysRevLett.117.110802,NatCommun.11.5382,PhysRevLett.124.020501,PhysRevLett.128.173602,Nat.548.187,Nat.576.65}, some researches also reported that sensors possessing unique NH characteristics have improved performance \cite{,PhysRevLett.123.180403,PhysRevRes.4.013113,PhysRevAppl.17.014034}.
While some studies suggested that the reported performance improvement might not have taken a full account of the effects of introduced noise \cite{PhysRevA.98.023805,NewJPhys.21.083002,NatCommun.11.1610,NatCommun.11.2454,ACSPhoton.9.1554}, other works also indicated that the improved performance could exist even considering noise \cite{NatCommun.9.4320,PhysRevLett.123.180501,Nat.607.697}.
A recent study proposed that when comparing the performance of quantum sensors, it is essential to fix the quantum resources consumed by these sensors \cite{PhysRevLett.131.160801}.
As far as we know, few studies have considered that, on the premise of the existence of superior sensing performance, the advantages of \textsl{non-Hermitian quantum sensing} (NHQS) could be restricted by certain factors in practical application, such as plenty of potential resource consumption with a low utilization rate.

In this paper, an investigation of the advantages and restrictions of \PT-symmetric QS is reported.
Based on two schemes of quantum simulations of \PT-symmetric systems, we verify that the EP indeed dramatically increases the population shift concerning the perturbation of estimated parameter, which seems to be exploited to improve the performance of QS.
However, we also find that the EP may not always enhance the susceptibility regarding the estimated parameter, and the non-Hermiticity may reduce the susceptibility even without involving the EP.
We further demonstrate that the advantages of NHQS over its counterpart \textsl{Hermitian quantum sensing} (HQS) deeply depend on whether the extra degrees of freedom introduced to construct \PT-symmetric systems carry additional quantum information.
Moreover, the restrictions on practically improving performance in NHQS mainly arise from the post-selection involving additional resource consumption in realistic constructions of \PT-symmetric systems.
While, within the framework of resource theory, EPQS indeed possesses potential superiority since the resource utilization rate can reach $100\%$ at periodic points.
This work has a potential promotion to develop quantum precision measurement.

The paper is organized as follows.
In Sec.~\ref{sec2}, two schemes of quantum simulations of the \PT-symmetric system based on the trapped-ion platform are briefly introduced.
In Sec.~\ref{sec3}, \PT-symmetric EPQS is verified by the population shift and susceptibility, and the advantages and restrictions of NHQS are investigated by the \textsl{quantum Fisher information} (QFI) and sensitivity bound.
Finally, conclusions are given in Sec.~\ref{sec4}.

\section{\label{sec2} \PT-symmetric quantum system}
\PT-symmetric quantum system, of which energy is balanced of gain ($|1\rangle$) and loss ($|2\rangle$), is governed by a NH Hamiltonian:
\begin{eqnarray}
\label{HPT}
H_\mathcal{PT}=\frac{\omega}{2}\sigma_x+\mathrm{i}\frac{\gamma}{2}\sigma_z,
\end{eqnarray}
where $\omega$ and $\gamma$ are the coupling rate and tunable gain-loss rate, $\sigma_{x,y,z}$ are Pauli matrixes with $\sigma_z=|1\rangle\langle1|-|2\rangle\langle2|$, and $\mathrm{E}_\pm=\pm\kappa/2$ with $\kappa=\sqrt{\omega^2-\gamma^2}$ are the eigenvalues of $H_\mathcal{PT}$ (the eigenvectors $|\mathrm{E}_\pm\rangle$ accord with the standard Dirac inner product).
The ratio $\gamma/\omega$ usually quantifies the non-Hermiticity in the system.
When $\gamma/\omega\in(0,1)$, $H_\mathcal{PT}$ is in the \PT-symmetric phase; and $\gamma/\omega=1$ is regarded as the EP of $H_\mathcal{PT}$ \cite{Sci.363.7709,PhysRevA.100.062131}; while $\gamma=0$ denotes an absence of non-Hermiticity, and the \PT-symmetric system degrades into a Rabi oscillator.
Generally, NH dynamics is non-unitary due to the non-Hermiticity in the system, the norm-preserving dynamics generated by a NH Hamiltonian $H_\mathrm{NH}=H_++H_-$ (with $H_\pm=\pm H_\pm^\dag$ and $H_-=-\mathrm{i}\Gamma$) \cite{PhysRevLett.109.230405} ($\hbar=1$) is
\begin{eqnarray}
\label{nHeq}
\dot{\rho}_t=-\mathrm{i}\left[H_+,\rho_t\right]-\left\{\Gamma,\rho_t\right\}+2\Tr\left(\rho_t\Gamma\right)\rho_t,
\end{eqnarray}
where $\rho_t$ is a normalized density operator.
The solution of Eq.~\eqref{nHeq} also can be simply expressed as
\begin{subequations}
\label{RP}
\begin{eqnarray}
\label{Rt1}
\rho_t=\frac{U_\mathcal{PT}\rho_0U_\mathcal{PT}^\dag}{\Tr\left(U_\mathcal{PT}\rho_0U_\mathcal{PT}^\dag\right)},
\end{eqnarray}
where $U_\mathcal{PT}=\exp(-\mathrm{i}H_\mathcal{PT}t)$.
Considering an initial pure state $\rho_0=|\psi_0\rangle\langle\psi_0|$, Eq.~\eqref{Rt1} is reduced to
\begin{eqnarray}
\label{Psit}
|\psi_t\rangle=\frac{U_\mathcal{PT}|\psi_0\rangle}{\sqrt{\langle\psi_0|U_\mathcal{PT}^\dag U_\mathcal{PT}|\psi_0\rangle}}=\mathrm{c_n}\cdot U_\mathcal{PT}|\psi_0\rangle,
\end{eqnarray}
\end{subequations}
with the normalized factor $\mathrm{c_n}=(\langle\psi_0|U_\mathcal{PT}^\dag U_\mathcal{PT}|\psi_0\rangle)^{-1/2}$.
Note that although the mathematical form of Eq.~\eqref{RP} is unitary-like, both $U_\mathcal{PT}$ and the corresponding NH dynamics are still non-unitary in the standard Dirac sense.
The major distinction between conventional (Hermitian) and \PT~quantum mechanics is the definition of inner product, hence the \PT~inner product, whose associated norm is positive definite, is introduced for making sense of \PT~quantum mechanics \cite{PhysRevLett.89.270401,RepProgPhys.70.947}.
To clear and concise, we set the experiment-dependent parameter $\omega\rightarrow1$ with the corresponding period of dynamics $T=2\pi/\omega\rightarrow2\pi$, and scale the evolution time with $\tau=\kappa t$ in subsequent numerical calculations and plots.

\begin{figure}[!htbp]
\centering
\includegraphics[width=0.48\textwidth]{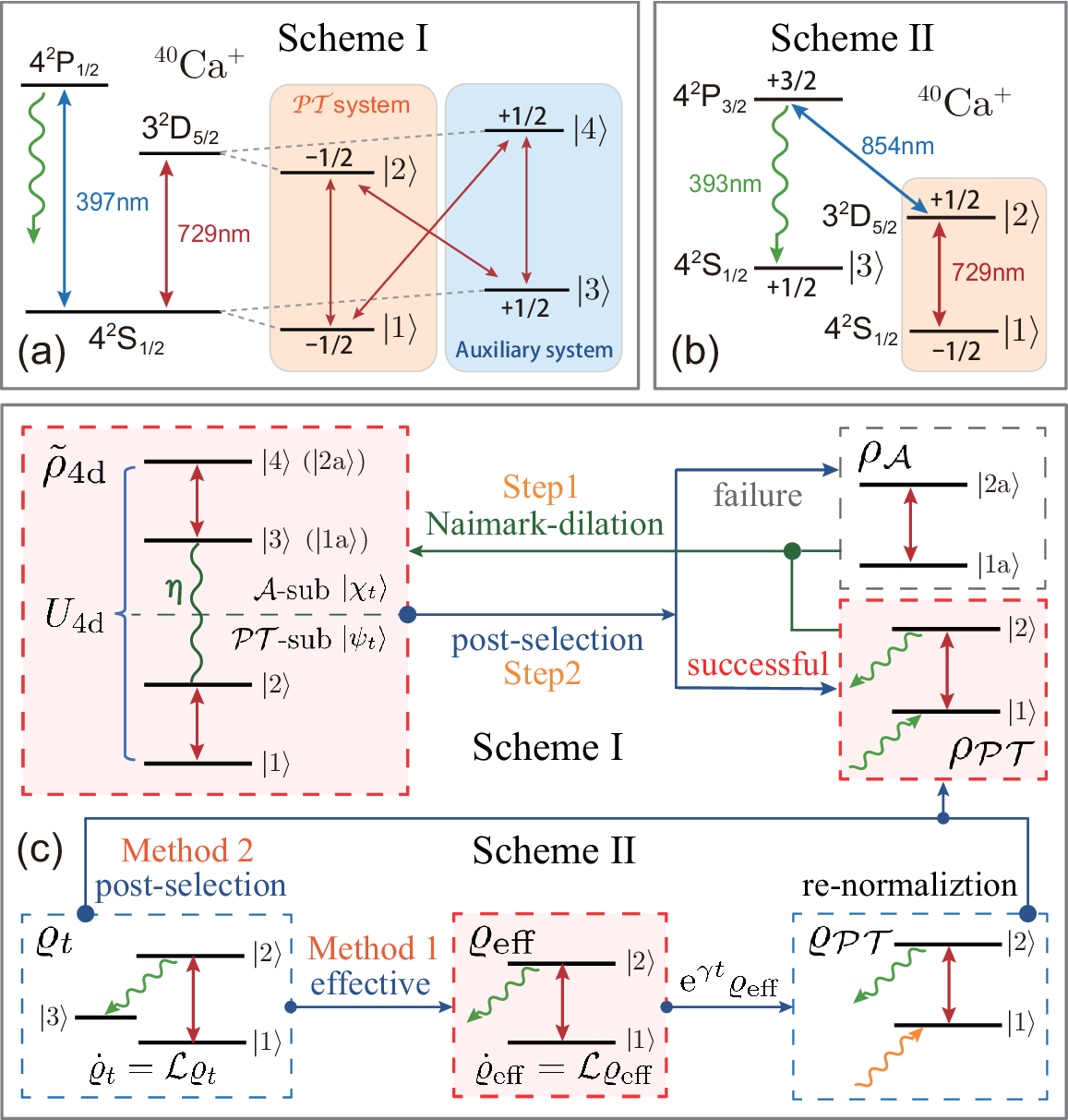}
\caption{(Color online)
Experimental schemes of quantum simulations of the \PT-symmetric system based on a trapped ion.\\
\textbf{(a)} \textit{Scheme I} with the digital quantum simulation method:
Choosing Zeeman sublevels of the trapped $^{40}\mathrm{Ca}^+$ ion as $|1\rangle=|4^2S_{1/2}(m_J=-1/2)\rangle$, $|2\rangle=|3^2D_{5/2}(m_J=-1/2)\rangle$, $|3\rangle=|4^2S_{1/2}(m_J=+1/2)\rangle$ and $|4\rangle=|3^2D_{5/2}(m_J=+1/2)\rangle$.
Any $S$-$D$ transition (red lines) can be implemented by a resonance drive using $729~nm$ laser.
Another $397~nm$ laser is used for cooling and fluorescence detection.
Initial state preparation and arbitrary unitary operation can be decomposed into appropriate sequences of $S$-$D$ equatorial rotation  $R(\theta,\phi)=\exp[-\mathrm{i}\theta(\cos\phi~\sigma^m_x+\sin\phi~\sigma^m_y)/2]$, where $\theta$ and $\phi$ are the rotation angle and laser phase, and $\sigma^m_{x,y}$ are Pauli matrixes.~\cite{Wu2023arXiv}
\textbf{(b)} \textit{Scheme II} with the analog quantum simulation method:
Choosing Zeeman sublevels of the trapped $^{40}\mathrm{Ca}^+$ ion as $|1\rangle=|4^{2}S_{1/2}(m_{J}=-1/2)\rangle$, $|2\rangle=|3^2D_{5/2}(m_J=+1/2)\rangle$ and $|3\rangle=|4^{2}S_{1/2}(m_{J}=+1/2)\rangle$.
The trapped $^{40}\mathrm{Ca}^+$ ion is initially prepared in $|1\rangle$, and then is resonantly driven to $|2\rangle$ by $729~nm$ laser, $854~nm$ laser induces a tunable loss in $|2\rangle$ through coupling $|2\rangle$ to a short-life level $|4^2P_{3/2}(m_{J}=+3/2)\rangle$ which quickly decays to $|3\rangle$.
Another $393~nm$ laser is used for cooling and fluorescence detection.~\cite{PhysRevA.103.L020201}
\textbf{(c)} Correlations between systems in this paper, details are given in Appendixes.
}
\label{fig:Ex}
\end{figure}

On the other hand, within the framework of standard quantum mechanics, it is quite a challenge to realize a NH Hamiltonian in a single quantum system.
Currently, the primary method to realize quantum simulations of \PT-symmetric Hamiltonians is through introducing extra degrees of freedom, such as an auxiliary system \cite{Sci.364.878,PhysRevLett.125.240506,PhysRevLett.123.230401,PhysRevLett.127.090501,PhysRevA.105.032405,Wu2023arXiv} (\textit{Scheme~I} in Fig~\ref{fig:Ex}~\blue{(a)}) or an external environment \cite{NatPhys.15.1232,PhysRevLett.126.083604,PhysRevA.103.L020201} (\textit{Scheme~II} in Fig~\ref{fig:Ex}~\blue{(b)}).
For the first one, based on the Naimark-dilation theory and the post-selection processing, a \PT-symmetric system can be reinterpreted as a subspace in a larger Hilbert space, and the \PT-symmetric non-unitary dynamics is equivalent to the unitary dynamics in enlarged Hilbert space.
For the other one, considering an open quantum system, whose dynamics obeys the Lindblad master equation, is dominated by an effective NH Hamiltonian when quantum jumps are ignored.
The experimental researches \cite{Nat.470.486,Wu2023arXiv,PhysRevA.103.L020201} have determined that the trapped-ion platform is feasible in realizing a \PT-symmetric system based on these schemes.
In the following, two schemes for quantum simulating \PT-symmetric system will be briefly introduced (more details see Appx.~\ref{apA} and \ref{apB}).
To facilitate reading, relevant systems in this paper with corresponding abbreviations and symbols are listed in Table~\ref{table:1}, and the correlations between these systems are shown in Fig.~\ref{fig:Ex}~\blue{(c)}.
\begin{table}[h]
\centering
\caption{Symbols and corresponding quantum systems.}
\label{table:1}
	\begin{tabular}{ll}
		\hline
		\hline
		System (Abbr.) & Symbols \\
		\hline
        \PT-symmetric system & $\rho_t$, $\rho_\mathcal{PT}$\\
        Enlarged Hermitian system & $\tilde{\rho}_\mathrm{4d}$, $|\Psi_t\rangle$\\
        Auxiliary system (\A~system) & $\rho_\mathcal{A}$\\
        \PT-symmetric subsystem (\PT-sub) & $|\psi_t\rangle$\\
        Auxiliary subsystem (\A-sub) & $|\chi_t\rangle$\\
        Dissipative two-level system & $\varrho_\mathrm{eff}$, $\rho_\mathrm{eff}$\\
        Three-level system & $\varrho_t$\\
        Artificial \PT-symmetric system & $\varrho_\mathcal{PT}$\\
		\hline
		\hline
	\end{tabular}
\end{table}

\begin{figure}[!htbp]
\centering
\includegraphics[width=0.48\textwidth]{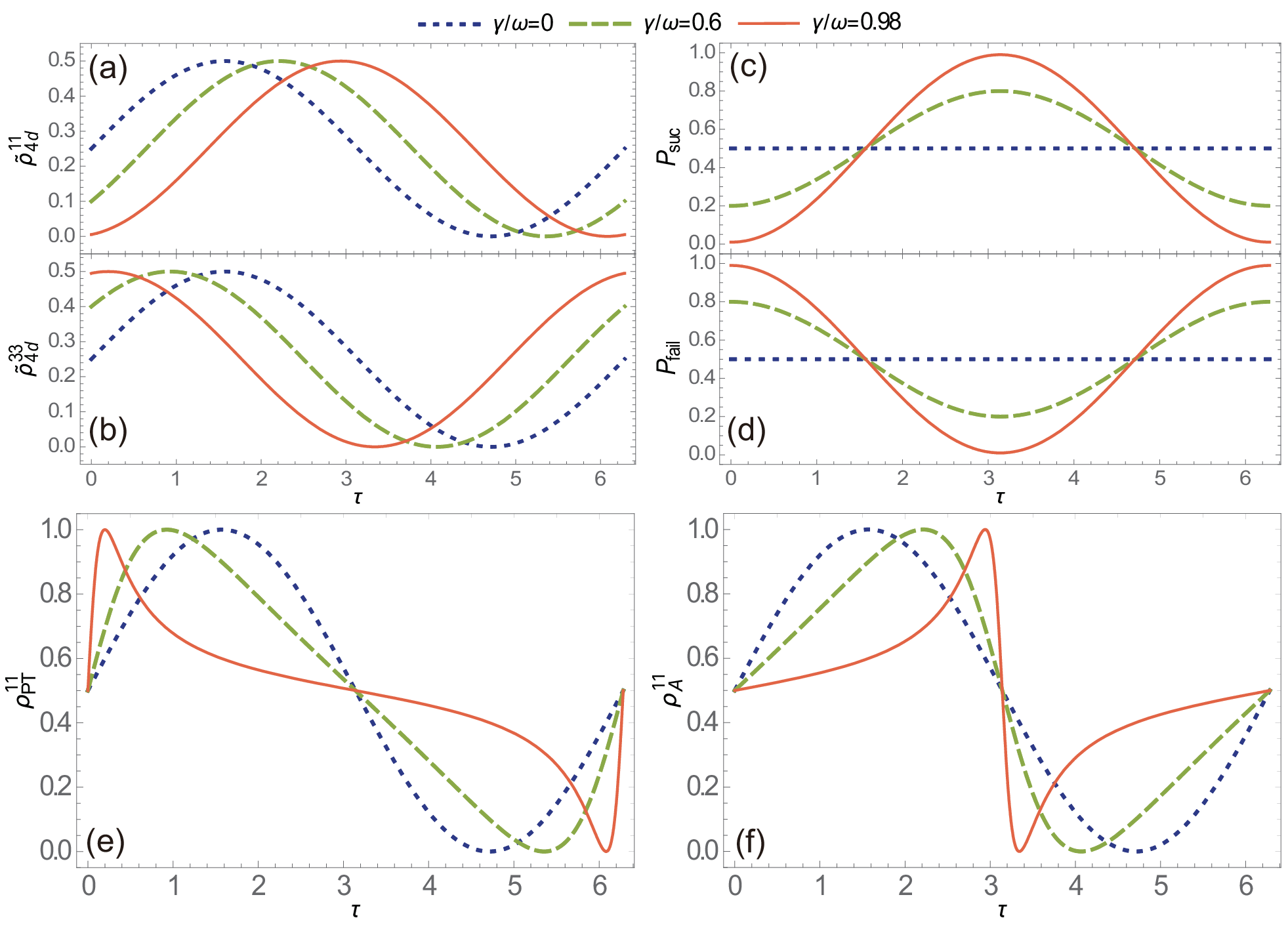}
\caption{(Color online)
Population and related quantities versus the scaled evolution time $\tau=\kappa t$ for different non-Hermiticity $\gamma/\omega$ with the initial condition $|\psi_0\rangle=|+\rangle_y$ and $|\Psi_0\rangle=(\psi_0~\eta\psi_0)^\mathrm{T}$.
Population of subsystems in enlarged Hermitian system before the post-selection:
\textbf{(a)} \PT-sub $\tilde{\rho}_\mathrm{4d}^{11}$;
\textbf{(b)} \A-sub $\tilde{\rho}_\mathrm{4d}^{33}$.
Related quantities of the post-selection:
\textbf{(c)} Success rate $p_\mathrm{suc}$;
\textbf{(d)} Failure rate $p_\mathrm{fail}$.
Population of systems after the post-selection:
\textbf{(e)} \PT-symmetric system $\rho_\mathcal{PT}^{11}$;
\textbf{(f)} \A~system $\rho_\mathcal{A}^{11}$.
}
\label{fig:4d}
\end{figure}

\textit{Scheme~I: Naimark-dilated quantum system.--}
Utilizing the Naimark-dilation theory, the \PT-symmetric system can be regarded as a NH subsystem embedded into a larger Hermitian system, which is achieved by embedding the \PT-symmetric subspace $\mathcal{H_{PT}}$ into a larger Hermitian space $\mathcal{H}_\mathrm{H(4d)}=\mathcal{H_{PT}}_{(2\mathrm{d})}\oplus\mathcal{H}_{\mathcal{A}(2\mathrm{d})}$ with an auxiliary subspace $\mathcal{H_A}$ \cite{PhysRevLett.101.230404} by constructing a Hermitian metric operator $\eta=\eta^\dag$ based on the pseudo-Hermiticity condition $\eta H_\mathcal{PT}=H_\mathcal{PT}^\dag\eta$ \cite{JMathPhys.43.2814}.
With the basis $\{|i\rangle\}$ ($i=1,2,3,4$), the \PT-symmetric subsystem (\PT-sub) $|\psi_t\rangle$ is denoted with $\{|1\rangle,|2\rangle\}$, and the auxiliary subsystem (\A-sub) $|\chi_t\rangle$ is denoted with $\{|3\rangle,|4\rangle\}$.
Obeying the \PT~inner product $(\mu,\nu)=(\mathcal{PT}\mu)\cdot\nu$ \cite{PhysRevLett.89.270401}, the eigenvectors $|E_\pm\rangle$ of $H_\mathcal{PT}$ are satisfied with $(E_\pm,E_\mp)=0$ and $(E_\pm,E_\pm)=\pm1$.
Arranging them as $\Phi=(|E_+\rangle~|E_-\rangle)$, the metric operator is defined by
\begin{eqnarray}
\label{eta}
\eta=\left(\Phi\Phi^\dagger\right)^{-1}=\frac{1}{\kappa}\left(\gamma\sigma_y+\mathbb{I}\right),
\end{eqnarray}
which is considered as a synchronization link of dynamics between \A-sub $|\chi_t\rangle$ and \PT-sub $|\psi_t\rangle$:
\begin{eqnarray}
\label{chi}
|\chi_t\rangle=\eta|\psi_t\rangle,
\end{eqnarray}
and the enlarged Hermitian system is constructed as
\begin{align}
\label{Psi}
|\Psi_t\rangle=\mathrm{C_n}\left(
\begin{bmatrix}
1\\
0\\
\end{bmatrix}
\otimes|\psi_t\rangle+
\begin{bmatrix}
0\\
1\\
\end{bmatrix}
\otimes|\chi_t\rangle\right)
=\mathrm{C_n}
\begin{pmatrix}
\psi_t\\
\chi_t\\
\end{pmatrix},
\end{align}
where $\mathrm{C_n}=1/\sqrt{\langle\psi_t|\psi_t\rangle+\langle\chi_t|\chi_t\rangle}$ is a normalized factor, and $\tilde{\rho}_\mathrm{4d}=|\Psi_t\rangle\langle\Psi_t|$ represents the normalized density operator of enlarged Hermitian system with the initial state $\tilde{\rho}_0=|\Psi_0\rangle\langle\Psi_0|$ related to $|\Psi_0\rangle=(\psi_0~\chi_0)^\mathrm{T}$, where $|\chi_0\rangle=\eta|\psi_0\rangle$ is the initial state of \A-sub and $|\psi_0\rangle$ for \PT-sub.
The corresponding time-evolution operator is
\begin{eqnarray}
\label{UH}
U_\mathrm{H}
=\begin{pmatrix}
U_\mathcal{PT}&0\\
0&\eta U_\mathcal{PT}\eta^{-1}\\
\end{pmatrix},
\end{eqnarray}
and then Eq.~\eqref{Psi} can be expressed in $|\Psi_t\rangle=U_\mathrm{4d}|\Psi_0\rangle$, where $U_\mathrm{4d}=(\mathrm{C_n}\mathrm{c_n})\cdot U_\mathrm{H}=\exp(-\mathrm{i}H_\mathrm{4d}t)$.
Now, with the Naimark-dilation theory, we have obtained the unitary time-evolution operator $U_\mathrm{4d}U_\mathrm{4d}^\dag=\mathbb{I}$ associated with the Hermitian Hamiltonian $H_\mathrm{4d}=H_\mathrm{4d}^\dag$, and the corresponding unitary time evolution $\tilde{\rho}_\mathrm{4d}=U_{\mathrm{4d}}\tilde{\rho}_{0}U_{\mathrm{4d}}^\dag$.
Population $P_{i(t)}=\tilde{\rho}_\mathrm{4d}^{ii}$ ($i=1,2,3,4$) of enlarged Hermitian system are exhibited in Figs.~\ref{fig:4d}~\blue{(a)~(b)}, where $\tilde{\rho}_\mathrm{4d}^{ii}$ are matrix diagonal elements of $\tilde{\rho}_\mathrm{4d}$.

Considering the post-selection, we can obtain the normalized population for \PT-sub:
\begin{subequations}
\label{P}
\begin{eqnarray}
\label{P1P2}
\tilde{P}_1=\frac{P_1}{P_1+P_2};~\tilde{P}_2=\frac{P_2}{P_1+P_2},
\end{eqnarray}
and the corresponding normalized population for \A-sub:
\begin{eqnarray}
\label{P3P4}
\tilde{P}_3=\frac{P_3}{P_3+P_4};~\tilde{P}_4=\frac{P_4}{P_3+P_4}.
\end{eqnarray}
\end{subequations}
If the post-selection is executed successfully, the enlarged Hermitian system with $P_{i(t)}=\tilde{\rho}_\mathrm{4d}^{ii}~(i=1,2)$ in Eq.~\eqref{P1P2} would collapse to the \PT-symmetric system with $p_{i(t)}=\rho_t^{ii}~(i=1,2)$, where $\rho_t^{ii}$ are matrix diagonal elements of $\rho_t$ in Eq.~\eqref{Rt1}.
The success rate and failure rate of post-selection are respectively denoted as $p_\mathrm{suc}=P_1+P_2$ and $p_\mathrm{fail}=1-p_\mathrm{suc}$ shown in Figs.~\ref{fig:4d}~\blue{(c)~(d)}.
After executing the post-selection, normalized density matrixes of \PT-symmetric system $\rho_\mathcal{PT}$ and auxiliary system (\A~system) $\rho_\mathcal{A}$ are respectively degraded into
\begin{subequations}
\label{4L}
\begin{eqnarray}
\label{4LPT}
\rho_\mathcal{PT}=\rho_t=\frac{|\psi_t\rangle\langle\psi_t|}{\Tr\left(|\psi_t\rangle\langle\psi_t|\right)}
=\begin{pmatrix}
\rho_t^{11}&\rho_t^{12}\\
\rho_t^{21}&\rho_t^{22}\\
\end{pmatrix},
\end{eqnarray}
where $\rho_t^{ij}~(i,j=1,2)$ with $|\psi_0\rangle=|+\rangle_y$, and
\begin{eqnarray}
\label{4LA}
\rho_\mathcal{A}=\frac{|\chi_t\rangle\langle\chi_t|}{\Tr\left(|\chi_t\rangle\langle\chi_t|\right)}
=\begin{pmatrix}
\rho_t^{22}&\rho_t^{21}\\
\rho_t^{12}&\rho_t^{11}\\
\end{pmatrix},
\end{eqnarray}
\end{subequations}
where $\rho_t^{ij}~(i,j=1,2)$ with $|\psi_0\rangle=|-\rangle_y$, and the pure states $|\pm\rangle_y=(|1\rangle\pm\mathrm{i}|2\rangle)/\sqrt{2}$ are the eigenvectors of the Pauli matrix $\sigma_y$.
Population of $\rho_\mathcal{PT}$ and $\rho_\mathcal{A}$ are respectively exhibited in Figs.~\ref{fig:4d}~\blue{(e)~(f)}.

Note that the \PT-symmetric system (or \A~system) obtained from executing the post-selection is different from the \PT-sub (or \A-sub) in the enlarged Hermitian system.
The former one possessing trace-preserving dynamics is obtained by executing the post-selection on the enlarged Hermitian system; while the latter one without performing the post-selection, of which dynamics is not trace-preserving.
Although the \PT-sub (or \A-sub) placed in the subspace of the larger Hermitian space possesses non-trace-preserving dynamics, the enlarged Hermitian system remains trace-preserving dynamics.
Additionally, Eq.~\eqref{4L} indicates that \A-sub indeed possesses the \PT-symmetry due to the synchronization link $\eta$ with \PT-sub, the EPs of \PT-sub and \A-sub are overlapping consequently.
The interlaced dynamics behavior shown in Figs.~\ref{fig:4d}~\blue{(e)~(f)} also demonstrates that two subsystems possess an anti-mirror-symmetric correlation, and the enlarged Hermitian system is combined by two synchronized \PT-symmetric subsystems (i.e., the enlarged Hermitian system can be regarded as a pseudo-dual-\PT-symmetric system).
More details of \textit{Scheme I} are given in Appx.~\ref{apA}.\\

\begin{figure}[!htbp]
\centering
\includegraphics[width=0.48\textwidth]{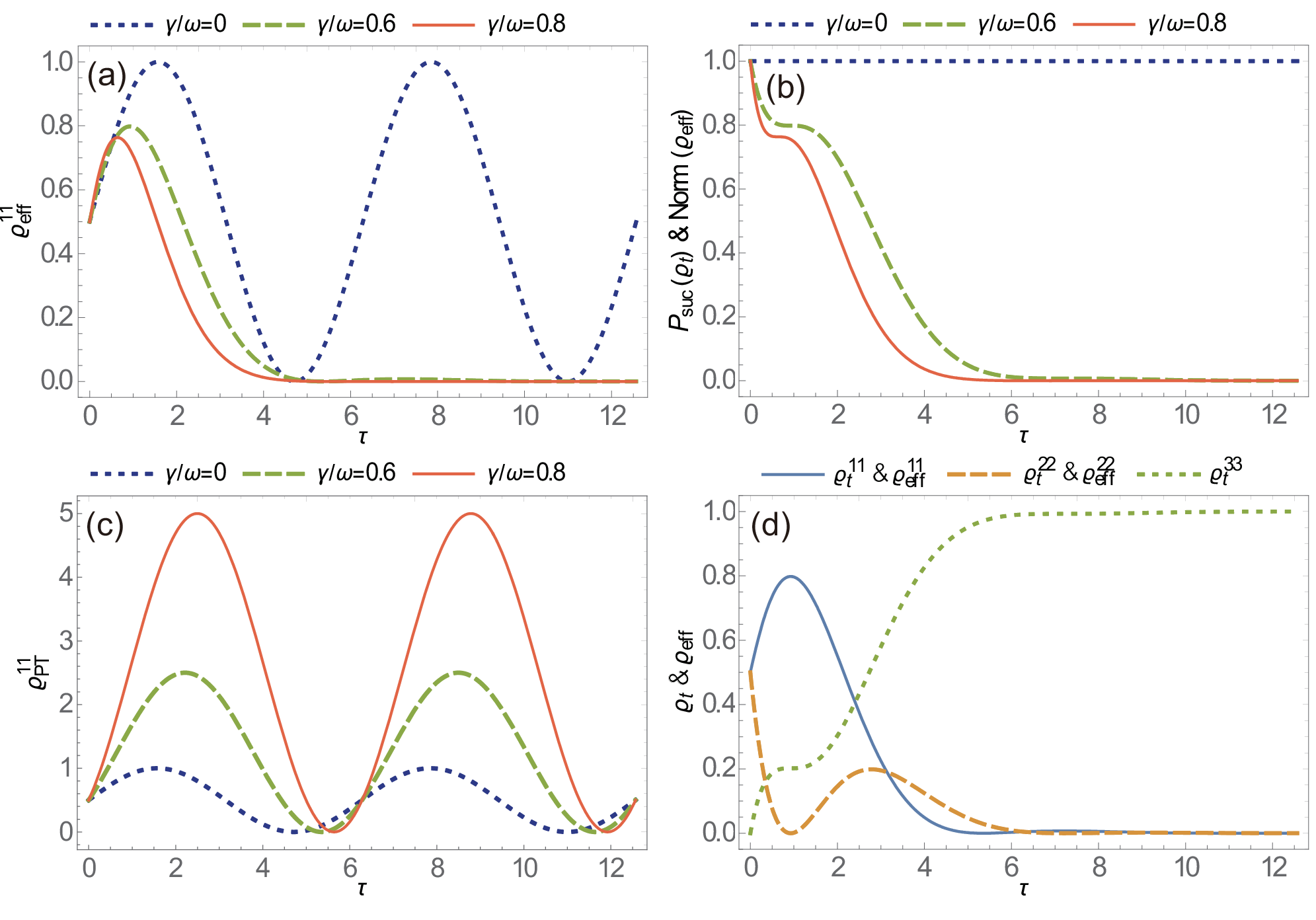}
\caption{(Color online)
Population and related quantities versus $\tau$ for different $\gamma/\omega$ with the initial condition $|\psi_0\rangle=|+\rangle_y$:
\textbf{(a)} Dissipative two-level system $\varrho_\mathrm{eff}^{11}$;
\textbf{(b)} Success-rate $p_\mathrm{suc}$ of the post-selection for three-level system $\varrho_t$ (i.e., norm of $\varrho_\mathrm{eff}$);
\textbf{(c)} Artificial \PT-symmetric system $\varrho_\mathcal{PT}^{11}$;
\textbf{(d)} Diagonal elements of density matrix (i.e., population of $|i\rangle$) of three-level system $\varrho_t$ and the corresponding dissipative two-level system $\varrho_\mathrm{eff}$ with $\gamma/\omega=0.6$.
}
\label{fig:3L}
\end{figure}

\textit{Scheme~II: Effective non-Hermitian Hamiltonian of open quantum system.--}
\PT-symmetry requires an exact energy balance of gain and loss, which is challenging to achieve in the quantum realm.
Coupling Hermitian systems with a dissipative reservoir can overcome this obstacle \cite{PhysRevLett.103.093902,Sci.346.328}.
Considering a dissipative two-level system (i.e., an open quantum system), whose energy is only with loss ($|2\rangle$) and without gain ($|1\rangle$), is described by an effective NH Hamiltonian with the basis $\{|i\rangle\}~(i=1,2)$:
\begin{eqnarray}
\label{Heff}
H_\mathrm{eff}=\frac{\omega}{2}\sigma_x-\mathrm{i}\gamma|2\rangle\langle2|=H_\mathcal{PT}-\mathrm{i}\frac{\gamma}{2}\mathbb{I},
\end{eqnarray}
where $H_\mathcal{PT}$ is given by Eq.~\eqref{HPT}.
The dissipative two-level system also can be expressed as a three-level system \cite{NatPhys.15.1232}, with the basis $\{|i\rangle\}~(i=1,2,3)$, the coherent transition is denoted by $|1\rangle\leftrightarrow|2\rangle$ with the coupling rate $\omega$, the loss of the system is represented by $|2\rangle\rightarrow|3\rangle$ with the tunable decay rate $\gamma$, and $|1\rangle\nleftrightarrow|3\rangle$.
The dynamics of the three-level system obeys the Lindblad master equation with a Liouvillian super-operator $\mathcal{L}$:
\begin{eqnarray}
\label{Leq}
\dot{\varrho}_t=\mathcal{L}\varrho_t=-\mathrm{i}\left[H_0,\varrho_t\right]+J\varrho_tJ^\dag-\frac{1}{2}\left\{J^\dag J,\varrho_t\right\},
\end{eqnarray}
where $H_0=\omega\sigma_x/2$ is the coherent transition Hamiltonian, and $J=\sqrt{\gamma}~|3\rangle\langle2|$ is the jump operator.
The dynamics of the dissipative two-level system governed by $H_\mathrm{eff}$ can be determined by a lower-dimension (3d $\rightarrow$ 2d) Lindblad master equation reduced from Eq.~\eqref{Leq}:
\begin{eqnarray}
\label{Lreff}
\dot{\varrho}_\mathrm{eff}=\mathcal{L}\varrho_\mathrm{eff}=-\mathrm{i}\left[H_\mathrm{eff},\varrho_\mathrm{eff}\right],
\end{eqnarray}
where $\varrho_\mathrm{eff}$ is a non-normalized density operator of the dissipative two-level system.
Eq.~\eqref{Lreff} indicates that the lower-dimension super-operator $\mathcal{L}$ without considering quantum jumps plays the same role as the effective NH Hamiltonian $H_\mathrm{eff}$ in the dissipative dynamics.
The solution of Eq.~\eqref{Lreff} also can be simply expressed as
\begin{eqnarray}
\label{reff}
\varrho_\mathrm{eff}=U_\mathrm{eff}\varrho_0U_\mathrm{eff}^\dag,
\end{eqnarray}
where $U_\mathrm{eff}=\exp(-\mathrm{i}H_\mathrm{eff}t)$, Eq.~\eqref{reff} can be reduced to $\varrho_\mathrm{eff}=|\psi_\mathrm{eff}\rangle\langle\psi_\mathrm{eff}|$ and $|\psi_\mathrm{eff}\rangle=U_\mathrm{eff}|\psi_0\rangle$ for initial pure states $\varrho_0=\rho_0=|\psi_0\rangle\langle\psi_0|$.
In the practical experiment \cite{PhysRevA.103.L020201}, the directly measured quantity is
\begin{eqnarray}
\label{rPT}
\varrho_\mathcal{PT}=\mathrm{e}^{\gamma t}\varrho_\text{eff},
\end{eqnarray}
where $\mathrm{e}^{\gamma t}$ denotes the gain term which is artificially added for simulating the characteristic of energy balance of gain and loss in the \PT-symmetric system, and $\varrho_\mathcal{PT}$ can be renormalized as
\begin{subequations}
\label{Reff}
\begin{align}
\label{Reff1}
\rho_\mathcal{PT}=\frac{\varrho_\mathcal{PT}}{\Tr(\varrho_\mathcal{PT})}=\frac{\mathrm{e}^{\gamma t}\varrho_\mathrm{eff}}{\Tr(\mathrm{e}^{\gamma t}\varrho_\mathrm{eff})}=\frac{\varrho_\mathrm{eff}}{\Tr(\varrho_\mathrm{eff})}=\rho_\mathrm{eff}.
\end{align}
Eq.~\eqref{Reff1} indicates that, after the re-normalization, both $\varrho_\mathcal{PT}$ and $\varrho_\mathrm{eff}$ degrade into the normalized $\rho_t=\rho_\mathcal{PT}=\rho_\mathrm{eff}$ in Eq.~\eqref{Rt1}, hence the effective NH Hamiltonian of the dissipative two-level system can be used to equivalently describe the \PT-symmetric system.
From the perspective of the post-selection, $\rho_\mathcal{PT}$ also can be obtained by directly performing the post-selection on $\varrho_t$:
\begin{align}
\label{Reff2}
\rho_\mathcal{PT}=\frac{1}{\varrho_t^{11}+\varrho_t^{22}}
\begin{pmatrix}
\varrho_t^{11}&\varrho_t^{12}\\
\varrho_t^{21}&\varrho_t^{22}\\
\end{pmatrix}
=\begin{pmatrix}
\rho_t^{11}&\rho_t^{12}\\
\rho_t^{21}&\rho_t^{22}\\
\end{pmatrix},
\end{align}
\end{subequations}
where $\varrho_t^{ij}~(i,j=1,2,3)$ are matrix elements of $\varrho_t$ given by solving Eq.~\eqref{Leq}, and $p_\mathrm{suc}=\varrho_t^{11}+\varrho_t^{22}$ is the success rate of post-selection, $\rho_t^{ij}~(i,j=1,2)$ are matrix elements of $\rho_t$ in Eq.~\eqref{Rt1}.
Population of artificial \PT-symmetric system $\varrho_\mathcal{PT}$ and dissipative two-level system $\varrho_\text{eff}$ with $|\psi_0\rangle=|+\rangle_y=(|1\rangle+\mathrm{i}|2\rangle)/\sqrt{2}$ (i.e., three-level system $\varrho_t$ with $|\tilde{\psi}_0\rangle=|\tilde{+}\rangle_y=(|1\rangle+\mathrm{i}|2\rangle+0|3\rangle)/\sqrt{2}$) are respectively exhibited in Figs.~\ref{fig:3L}~\blue{(c)~(d)}.

After executing the re-normalization or post-selection, the dissipative dynamics in Fig.~\ref{fig:3L}~\blue{(a)} would degrade into the \PT-symmetric dynamics in Fig.~\ref{fig:4d}~\blue{(e)}, which shows that constructing the \PT-symmetric system by introducing an environment also can be regarded as a kind of method of expanding dimension (e.g., 2d $\rightarrow$ 3d).
However, the steady state is associated with a vanishing success rate shown in Figs.~\ref{fig:3L}~\blue{(b)~(d)}, which indicates that, under a prolonged evolution, such an expansion will be broken.
As we all know, the Liouvillian super-operator $\mathcal{L}$ expressed by a NH matrix also possesses EPs which can be defined as the degeneracy points of $\mathcal{L}$ matrix \cite{PhysRevA.100.062131}.
It can be easy to determine that, at $\gamma/\omega=1$, the \textsl{Hamiltonian exceptional point} (HEP) of $H_{\mathcal{PT}}$ in Eq.~\eqref{HPT} is coincident with the \textsl{Liouvillian exceptional point} (LEP) of $\mathcal{L}$ matrix in Eq.~\eqref{Lreff}.
More details of \textit{Scheme II} are given in Appx.~\ref{apB}.

\section{\label{sec3} \PT-symmetric quantum sensing}
In this section, based on the two schemes mentioned above, in Sec.~\ref{sec3A}, we first examine (i) population shift and (ii) susceptibility, to verify whether the EP can potentially enhance QS.
And then, in Sec.~\ref{sec3B}, we discuss (iii) the condition for existing advantages of NHQS over HQS by using QFI and sensitivity bound, and investigate (iv) whether the post-selection imposes restrictions on improving the performance of NHQS.

\subsection{\label{sec3A} \PT-symmetric EP-enhanced quantum sensing}
Quantum sensing achieves ultra-precision sensing by using the unique quantum properties of \PT-symmetric systems, such as the eigenvectors degeneracy and the susceptibility divergence at the EP, which is called EPQS.\\

\begin{figure*}[!htbp]
\centering
\includegraphics[width=0.88\textwidth]{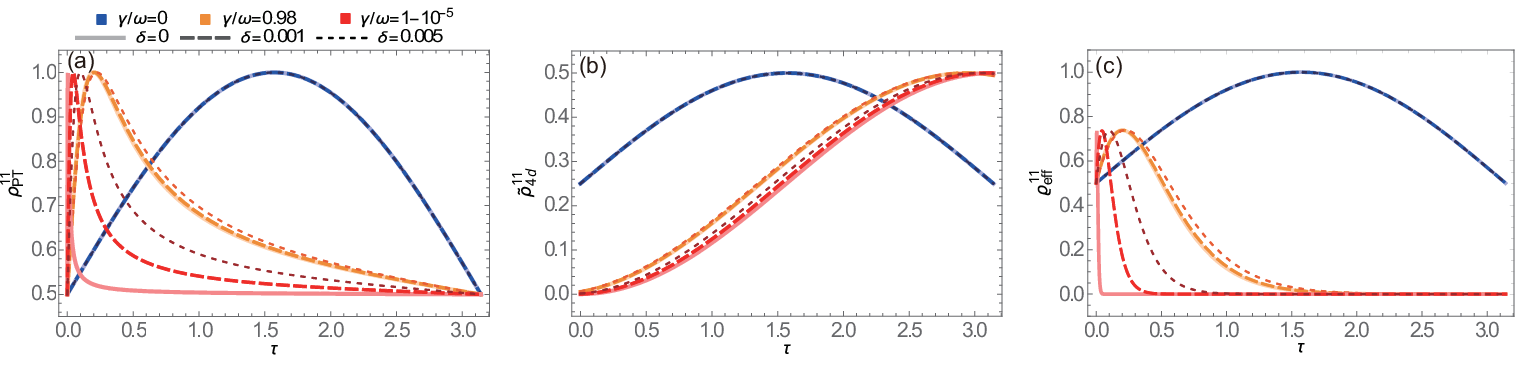}
\caption{(Color online)
Population shift with respect to the perturbation of estimated parameter $\delta$ versus $\tau$ for different $\gamma/\omega$:
\textbf{(a)} \PT-symmetric system $\rho_\mathcal{PT}$ (i.e., $\rho_t$ in Eq.~\eqref{Rt1});
\textbf{(b)} Enlarged Hermitian system $\tilde{\rho}_\mathrm{4d}$;
\textbf{(c)} Dissipative two-level system $\rho_\mathrm{eff}$.
}
\label{fig:4}
\end{figure*}
(i) \textit{Population shift with respect to the perturbation of estimated parameter.--}
Population shift is considered as a response to the perturbation of estimated parameter and directly reflects the ability to distinguish population change under the  perturbation.
From the perspective of quantum parameter estimation, the estimated parameter is directly coupled to the degrees of freedom of the system, the introduced parameter-dependent perturbation can be represented by a Hermitian Hamiltonian
\begin{eqnarray}
\label{Hper}
H_\delta=\frac{\delta}{2}\sigma_x.
\end{eqnarray}
Generally, the perturbation amplitude $\delta$ is rather weak, and the parameter-dependent \PT-symmetric Hamiltonian under the perturbation is
\begin{eqnarray}
\label{HPT2}
\tilde{H}_{\mathcal{PT}(\omega)}=H_\mathcal{PT}+H_\delta=\frac{\Omega}{2}\sigma_x+\mathrm{i}\frac{\gamma}{2}\sigma_z,
\end{eqnarray}
with $\Omega=\omega+\delta$, Eqs.~\eqref{HPT} and \eqref{Heff} are accordingly modified as $\omega\rightarrow\Omega$ and $H_\mathcal{PT}\rightarrow\tilde{H}_{\mathcal{PT}}$, the coupling rate $\omega$ is the parameter to be estimated.
Population shift of $|1\rangle$ related to the estimated-parameter perturbation $\delta$ for different systems are shown in Fig.~\ref{fig:4}.
It can be observed that, in the vicinity of the EP ($\gamma/\omega=1-10^{-5}$), even a tiny perturbation ($\delta/\omega=0.1\%$ and $0.5\%$) can lead to a significant population shift depicted by red curves, which indicates that the \PT-symmetric system (before introducing extra degrees of freedom) and the systems of two schemes (after introducing auxiliary system or external environment) are all extremely sensitive to the estimated-parameter perturbation near the EP.
Based on this, the EP appears the potential to enhance QS.
The main mechanism of the EP-enhanced response of perturbation is the strong dependency of energy level splitting on the parameter near the EP \cite{RevModPhys.93.015005}.
The derivative of eigenvalues and eigenvectors with respect to the estimated parameter diverges at EP, which is also one of the fundamental distinctions between NH and Hermitian Hamiltonians.
However, eigenvalue variations under the perturbation may not be an applicable measure of the comprehensive sensing performance at the EP \cite{ACSPhoton.9.1554}.
For further studying the EP-enhanced response, the susceptibility to the estimated parameter will be considered.\\

\begin{figure*}[!htbp]
\centering
\includegraphics[width=0.88\textwidth]{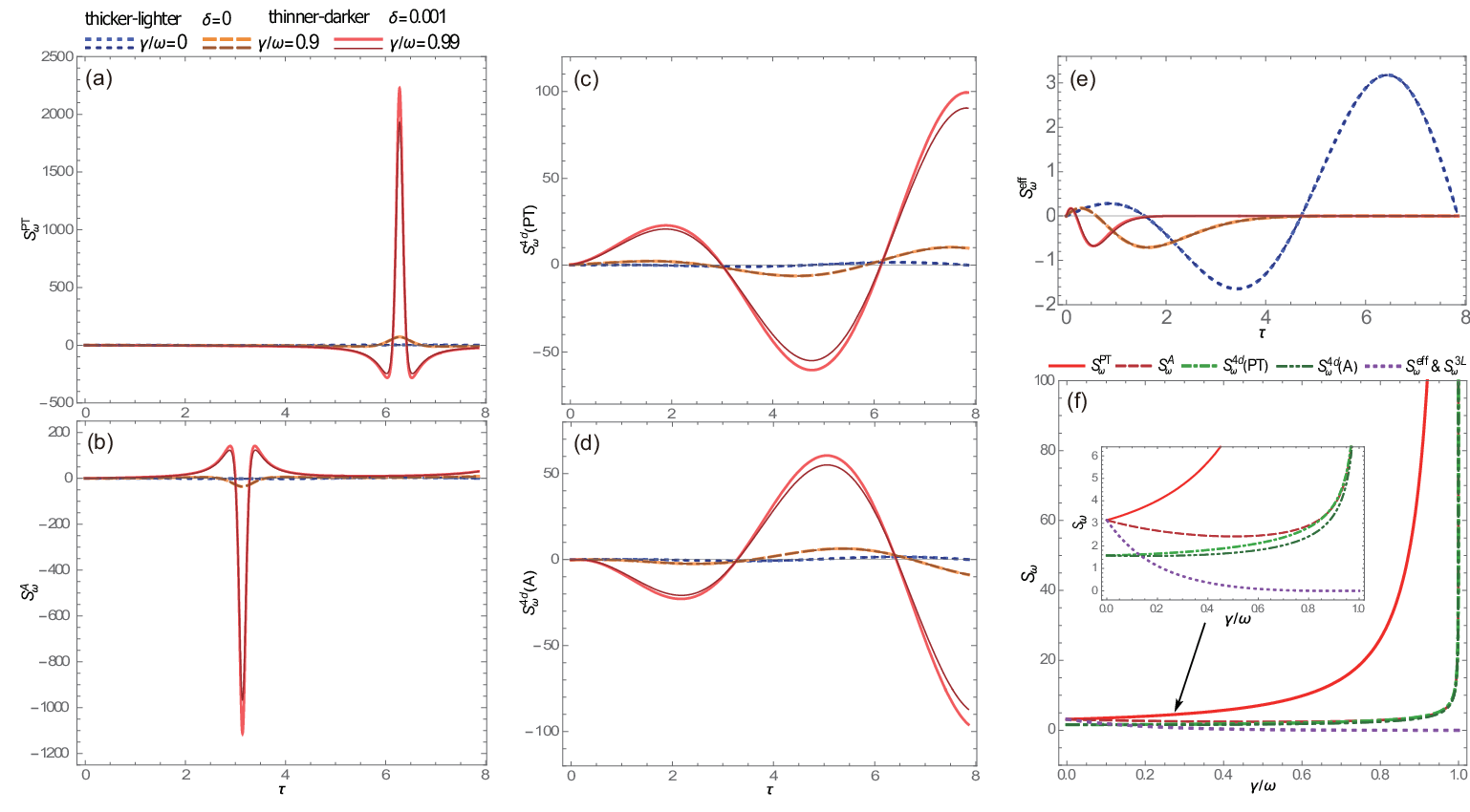}
\caption{(Color online)
Susceptibility $S_\omega$ in Eq.~\eqref{S_omega} with respect to the estimated parameter $\omega$ versus $\tau$ for different $\gamma/\omega$ and the perturbation $\delta$.
(a)-(d) for \textit{Scheme I}:
\textbf{(a)} \PT-symmetric system $\rho_\mathcal{PT}$, $S_\omega^\mathcal{PT}=\partial_\omega\rho_\mathcal{PT}^{11}$;
\textbf{(b)} \A~system $\rho_\mathcal{A}$, $S_\omega^\mathcal{A}=\partial_\omega\rho_\mathcal{A}^{11}$;
\textbf{(c)} \PT-sub (represented equivalently to enlarged Hermitian system with $\tilde{\rho}_\mathrm{4d}^{11}$ in this sense), $S_\omega^\mathrm{4d}(\mathcal{PT})=\partial_\omega\tilde{\rho}_\mathrm{4d}^{11}$;
\textbf{(d)} \A-sub, $S_\omega^\mathrm{4d}(\mathcal{A})=\partial_\omega\tilde{\rho}_\mathrm{4d}^{33}$.
\textbf{(e)} for \textit{Scheme II}: Dissipative two-level system  $\varrho_\mathrm{eff}$, $S_\omega^\mathrm{eff}=\partial_\omega\varrho_\mathrm{eff}^{11}$ (i.e., three-level system $\varrho_t^{11}=\varrho_\mathrm{eff}^{11}$, $S_\omega^\mathrm{3L}=\partial_\omega\varrho_t^{11}$).
\textbf{(f)} Susceptibility $S_\omega=\partial_\omega\rho_\omega^{ii}$ versus $\gamma/\omega$ for different parameterized states $\rho_\omega$ at $\tau=2\pi$.
}
\label{fig:5}
\end{figure*}
(ii) \textit{Susceptibility with respect to the estimated parameter.--}
Susceptibility quantifies how sensitive the population would respond to the variation of estimated parameter, and is directly defined by taking the derivative of population with respect to $\omega$:
\begin{eqnarray}
\label{S_omega}
S_\omega=\partial_\omega\rho_\omega^{ii},
\end{eqnarray}
where $\rho_\omega^{ii}$ is the $i$th diagonal element of the corresponding parameter-dependent density matrix $\rho_\omega$.
Usually, the maximum susceptibility corresponds to the optimal measurement point.

For \textit{Scheme I}, $S_\omega^\mathcal{PT}$ of \PT-symmetric system shown in Fig.~\ref{fig:5}~\textcolor{blue}{(a)} exhibits a divergent feature near the EP (red curves) and possesses a sensitive response to the perturbation (thinner curves).
It can be anticipated that the EP divergence of susceptibility appears the potential to achieve sensing with arbitrary precision.
And $S_\omega^\mathcal{A}$ of \A~system shown in Fig.~\ref{fig:5}~\blue{(b)} is consistent with $S_\omega^\mathcal{PT}$, also exists a divergent feature and sensitive response near the EP.
From the perspective of the enlarged Hermitian system, Figs.~\ref{fig:5}~\blue{(c)~(d)} show that, although $S_\omega^\mathrm{4d}$ exhibits a similar pattern of EP divergence and the sensitive response as that of the \PT-symmetric system, both the speed and intensity of EP divergence are noticeably restrained and weakened.
This primarily results from two systems connected by the synchronization link $\eta$ possessing interlaced evolution behavior of $S_\omega$ (in Figs.~\ref{fig:5}~\blue{(a)~(b)}), which leads to the internal constraint between \PT-sub and \A-sub in enlarged Hermitian system, weakening the EP divergence.

As for \textit{Scheme II}, $S_\omega^\mathrm{eff}$ of dissipative two-level system does not diverge near the EP shown in Fig.~\ref{fig:5}~\blue{(e)}, which is entirely different from that of the \PT-symmetric system.
Instead, $S_\omega^\mathrm{eff}$ approaches zero during a prolonged evolution and shows little response to the perturbation (thinner curves).
Particularly, $S_\omega^\mathrm{eff}$ of dissipative two-level system ($\gamma\neq0$) is distinctly weaker compared to its Hermitian counterpart ($\gamma=0$, blue-dotted curves in Fig.~\ref{fig:5}~\blue{(e)}).
In other words, the stronger the non-Hermiticity, the weaker $S_\omega^\mathrm{eff}$ (violet-dotted curves in Fig.~\ref{fig:5}~\blue{(f)}).
This contrasts with the superior feature observed in \textit{Scheme I}, in which non-Hermiticity can dramatically enhance the susceptibility (red and green curves in Fig.~\ref{fig:5}~\blue{(f)}).

Consequently, the anticipation that the EP divergence of susceptibility can be used to achieve sensing with arbitrary precision is not necessarily accurate, since the EP may not always enhance the susceptibility with respect to the estimated parameter.
On the contrary, the non-Hermiticity may reduce the susceptibility even without involving the EP.
To explore the condition of existing the superiority of NHQS, we would further investigate QFI and sensitivity bound in quantum measurements.

\subsection{\label{sec3B} Advantages and restrictions of non-Hermitian quantum sensing}
As mentioned above, the debate on whether NHQS is superior to HQS remains ongoing.
Here, a question is raised: What is the condition under which the superiority of NHQS exists?
Supposing that NHQS indeed possesses superior sensing performance, it is necessary to consider whether the performance improvement comes with potential costs or losses.\\

(iii) \textit{Sensitivity of non-Hermitian quantum sensors.--}
In quantum metrology, the sensitivity is represented by QFI which is defined as the supremum of classical Fisher information and regarded as a measure of the precision of parameter estimation.
The inverse of QFI sets the lower bound on the error of estimation \cite{JPhysA.53.023001}.
In NH or open quantum systems, parameterized processes typically involve non-unitary dynamics which can be mapped from unitary dynamics in larger systems, enabling the direct application of concepts such as QFI and sensitivity bound to parameter estimation based on such mapped dynamics \cite{PhysRevLett.131.160801}.
With the \textsl{symmetric logarithmic derivative} (SLD):
\begin{eqnarray}
\label{SLD}
\partial_\omega\rho_\omega=\frac{1}{2}\left(L_\omega\rho_\omega+\rho_\omega L_\omega\right),
\end{eqnarray}
the definition of QFI is given by
\begin{subequations}
\label{F}
\begin{eqnarray}
\label{F1}
\mathcal{F}_\omega=\Tr\left(\rho_\omega L_\omega^2\right),
\end{eqnarray}
where $L_\omega$ is SLD operator, $\rho_\omega$ is an arbitrary parameterized state.
With the proper orthogonal decomposition $\rho_\omega=\sum_n\varepsilon_n|\psi_n\rangle\langle\psi_n|$, Eq.~\eqref{F1} is expressed as
\begin{align}
\label{F2}
\nonumber\mathcal{F}_\omega=&\sum_n\varepsilon_n^{-1}(\partial_\omega\varepsilon_n)^2+\sum_n4\varepsilon_n\langle\partial_\omega\psi_n|\partial_\omega\psi_n\rangle\\
&-\sum_{n\neq m}8\varepsilon_n\varepsilon_m(\varepsilon_n+\varepsilon_m)^{-1}|\langle\partial_\omega\psi_n|\psi_m\rangle|^2.
\end{align}
For a pure parameterized state $\rho_\omega=|\psi_\omega\rangle\langle\psi_\omega|$, Eq.~\eqref{F2} can be reduced to $\mathcal{F}_\omega=4(\langle\partial_\omega\psi_\omega|\partial_\omega\psi_\omega\rangle-|\langle\psi_\omega|\partial_\omega\psi_\omega\rangle|^{2})$.
Considering a general two-level system \cite{PhysRevA.87.022337,PhysRevA.91.033805}, the expression of QFI can be explicitly obtained as
\begin{align}
\label{F3}
\mathcal{F}_\omega=\Tr\left[\left(\partial_\omega\rho_\omega\right)^2\right]+\frac{1}{\det\left(\rho_\omega\right)}\Tr\left[\left(\rho_\omega\partial_\omega\rho_\omega\right)^2\right],
\end{align}
\end{subequations}
and Eq.~\eqref{F3} also can be reduced to $\mathcal{F}_\omega=2\Tr[(\partial_\omega\rho_\omega)^2]$ for a pure $\rho_\omega$.
The channel QFI corresponding to the maximum achievable QFI is obtained by optimizing over all possible probe states.
The optimal probe (initial) state has been verified to always be a pure state due to the convexity of QFI \cite{PhysRevLett.123.250502}.
The inverse of channel QFI quantifies the sensitivity bound of parameter estimation:
\begin{eqnarray}
\label{CRB_F}
\delta\omega\geqslant\frac{1}{\sqrt{N\mathcal{F}_\omega}},
\end{eqnarray}
where $N$ is the repetitions of measurement, and Eq.~\eqref{CRB_F} is the well-known quantum Cram\`{e}r-Rao bound.

\begin{figure*}[!htbp]
\centering
\includegraphics[width=0.98\textwidth]{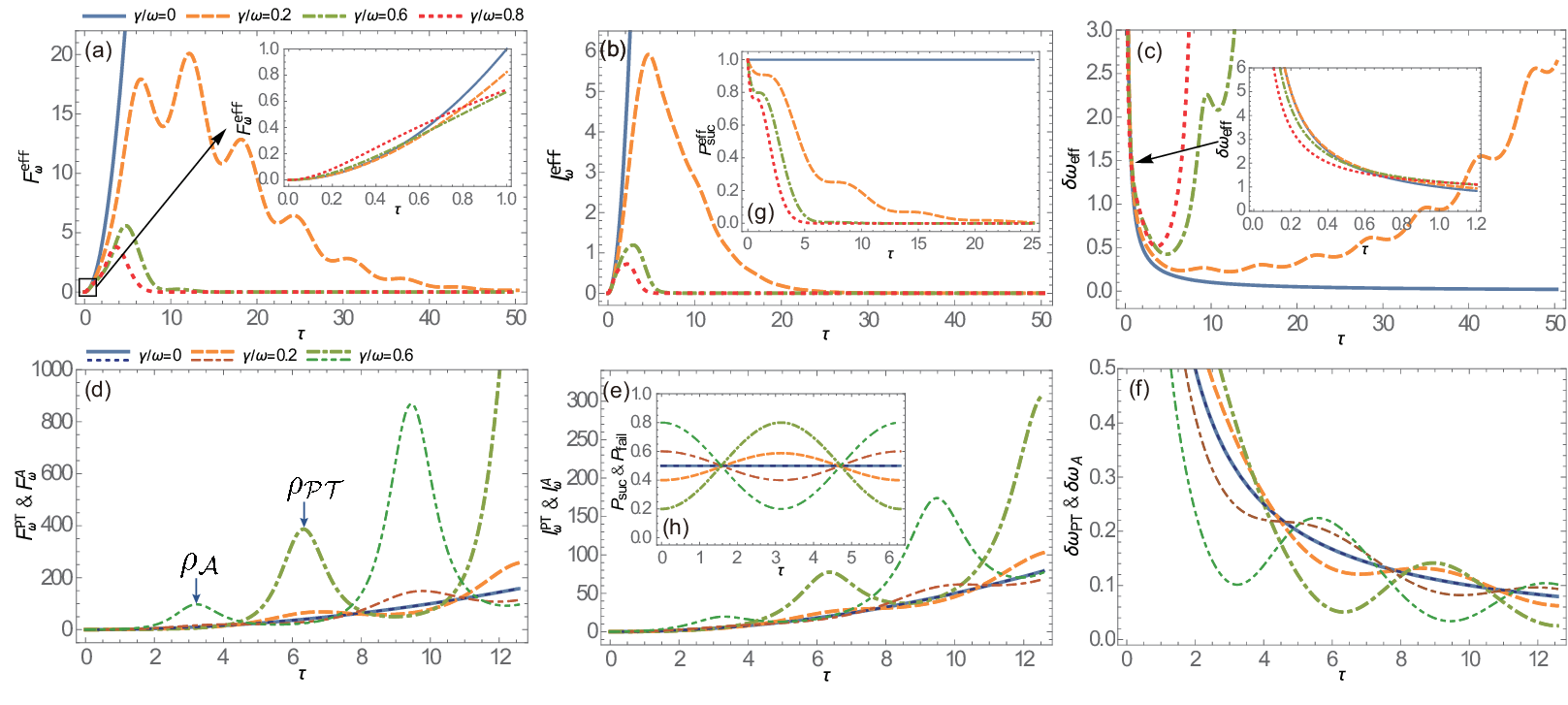}
\caption{(Color online)
QFI $\mathcal{F}_\omega$ in Eq.~\eqref{F}, $\mathcal{I}_\omega$ in Eqs.~\eqref{I4L} and \eqref{I3L}, and sensitivity bound $\delta\omega$ in Eq.~\eqref{CRB_F} with respect to the estimated parameter $\omega$ versus $\tau$ for different $\gamma/\omega$ with the averaged-measure scale $N=1$.
(a)-(c) for \textit{Scheme II}, dissipative two-level system $\varrho_\mathrm{eff}$:
\textbf{(a)} QFI with single-successful measurement, $\mathcal{F}_\omega^\mathrm{eff}$;
\textbf{(b)} QFI with repeated-averaged measurement, $\mathcal{I}_\omega^\mathrm{eff}$;
\textbf{(c)} Sensitivity bound with single-successful measurement, $\delta\omega_{\mathrm{eff}}$.
(d)-(f) for \textit{Scheme I}, \PT-symmetric system $\rho_\mathcal{PT}$ (thicker-lighter curves) and \A~system $\rho_\mathcal{A}$ (thinner-darker curves) of enlarged Hermitian system under the post-selection:
\textbf{(d)} QFI with single measurement, $\mathcal{F}_\omega^\mathcal{PT}$ and $\mathcal{F}_\omega^\mathcal{A}$;
\textbf{(e)} QFI with repeated-averaged measurement, $\mathcal{I}_\omega^\mathcal{PT}$ and $\mathcal{I}_\omega^\mathcal{A}$;
\textbf{(f)} Sensitivity bound with single measurement, $\delta\omega_\mathcal{PT}$ and $\delta\omega_\mathcal{A}$.
Other related quantities with the same system parameters:
\textbf{(g)} Success rate $p_\mathrm{suc}^\mathrm{eff}$ of the post-selection (i.e., norm of $\varrho_\mathrm{eff}$) in dissipative two-level system for \textit{Scheme II};
\textbf{(h)} Success rate $p_\mathrm{suc}$ (thicker-lighter curves) and failure rate $p_\mathrm{fail}$ (thinner-darker curves) of the post-selection in enlarged Hermitian system for \textit{Scheme I}.
}
\label{fig:6}
\end{figure*}

Note that realistic constructions of \PT-symmetric systems necessitate a post-selection involving a success rate, as a consequence, Eq.~\eqref{F} characterizes the estimated-parameter information obtained from a single measurement, and Eq.~\eqref{CRB_F} represents the corresponding sensitivity bound ($N=1$).
When performing repeated measurements, it is necessary to consider the success rate of post-selection.
By averaging the outcomes of repeated measurements, the information (i.e., QFI) obtained from a repeated-averaged measurement can be expressed.
As for \textit{Scheme I}, under the post-selection, the weighted QFI $\mathcal{I}_\omega$ obtained from a repeated-averaged measurement for the \PT-symmetric system and \A~system are given as
\begin{subequations}
\label{I4L}
\begin{eqnarray}
\label{I4LPT}
\mathcal{I}_\omega^\mathrm{suc}=\mathcal{F}_\omega^\mathcal{PT}\cdot p_\mathrm{suc},
\end{eqnarray}
\begin{eqnarray}
\label{I4LA}
\mathcal{I}_\omega^\mathrm{fail}=\mathcal{F}_\omega^\mathcal{A}\cdot p_\mathrm{fail},
\end{eqnarray}
\end{subequations}
where $\mathcal{I}_\omega^\mathrm{suc}$ represents QFI when the post-selection is executed successfully, the other $\mathcal{I}_\omega^\mathrm{fail}$ for failure.
By summing $\mathcal{I}_\omega^\mathrm{suc}$ and $\mathcal{I}_\omega^\mathrm{fail}$, the total QFI obtained from a repeated-averaged measurement involved the success rate is
\begin{subequations}
\label{I4Lpost}
\begin{eqnarray}
\label{I4Lsub}
\mathcal{I}_\omega^\mathrm{subs}=\mathcal{I}_\omega^\mathrm{suc}+\mathcal{I}_\omega^\mathrm{fail}.
\end{eqnarray}
Without (or before) the post-selection, the total QFI of enlarged Hermitian system uninvolved with the success rate is directly represented by $\mathcal{F}_\omega$ in Eq.~\eqref{F}:
\begin{eqnarray}
\label{I4Ltot}
\mathcal{I}_\omega^\mathrm{4d}=\mathcal{F}_\omega^\mathrm{4d}.
\end{eqnarray}
\end{subequations}
For \textit{Scheme II}, the total QFI of dissipative two-level system obtained from a repeated-averaged measurement is
\begin{eqnarray}
\label{I3L}
\mathcal{I}_\omega^\mathrm{eff}=\mathcal{F}_\omega^\mathrm{eff}\cdot p_\mathrm{suc}^\mathrm{eff},
\end{eqnarray}
where the success rate $p_\mathrm{suc}^\mathrm{eff}=\varrho_t^{11}+\varrho_t^{22}$ denotes the norm of $\varrho_\mathrm{eff}$.
The sensitivity bound represented by Eq.~\eqref{CRB_F} is correspondingly modified with $\mathcal{F}_\omega\rightarrow\mathcal{I}_\omega$:
\begin{align}
\label{CRB_I}
\delta\omega\geqslant\frac{1}{\sqrt{N \mathcal{I}_\omega}}.
\end{align}
For determining whether NHQS is superior to HQS, QFI $\mathcal{F}_\omega$, $\mathcal{I}_\omega$ and sensitivity bound $\delta\omega$ of the two schemes are exhibited in Fig.~\ref{fig:6} with the optimal probe state $|\psi_0\rangle=|+\rangle_y$.
Since the repetitions $N$ in $\delta\omega$ of a NH system is the same as its Hermitian counterpart, we plot subsequent figures with the averaged-measure scale $N=1$.

For \textit{Scheme II}, Figs.~\ref{fig:6}~\blue{(a)~(c)} indicate that $\mathcal{F}_\omega^\mathrm{eff}$ and $\delta\omega_\mathrm{eff}$ of the dissipative two-level system ($\gamma\neq0$) obtained from a single-successful measurement are generally lower compared to Hermitian counterparts ($\gamma=0$, blue-solid curves), and slightly surpass Hermitian counterparts during the initial period only (shown in subfigures).
The oscillatory evolution behavior of $\mathcal{F}_\omega^\mathrm{eff}$ can be attributed to the simultaneous encoding and loss of estimated-parameter information in the parameterized process.
And $\mathcal{F}_\omega^\mathrm{eff}$ completely loses during prolonged evolution due to the steady state does not contain any estimated-parameter information.
This loss results from the non-Hermiticity (i.e., dissipation) in the dissipative two-level system jointed with the invariance of total estimated-parameter information in the parameterized process.
In Fig.~\ref{fig:6}~\blue{(b)}, $\mathcal{I}_\omega^\mathrm{eff}$ involving the post-selection appears to be a notable decrease and accelerated loss compared to $\mathcal{F}_\omega^\mathrm{eff}$, because the success rate gradually vanishes as $p_\mathrm{suc}^\mathrm{eff}\rightarrow0$ shown in Fig.~\ref{fig:6}~\blue{(g)}.

As for \textit{Scheme I}, the synchronization link $\eta$ connects subsystems of enlarged Hermitian system and leads to the relevance of interlaced dynamics behaviors shown in Figs.~\ref{fig:4d}~\blue{(e)~(f)} between \PT-symmetric system and \A~system even after the post-selection.
This relevance makes evolution behaviors of $\mathcal{F}_\omega^\mathcal{PT}$ and $\mathcal{F}_\omega^\mathcal{A}$ obtained from a single measurement also exhibit an interlaced pattern shown in Fig.~\ref{fig:6}~\blue{(d)}.
Obviously, the introduced degrees of freedom of \A~system equally carry the estimated-parameter information.
Contrary to \textit{Scheme II}, the presence of non-Hermiticity ($\gamma\neq0$) based on \textit{Scheme I} significantly enhances $\mathcal{F}_\omega$ and $\delta\omega$ compared to Hermitian counterparts ($\gamma=0$, blue-solid curves).
Considering a repeated-averaged measurement, Figs.~\ref{fig:6}~\blue{(e)~(f)} also show that, despite the success rate of post-selection being oscillatory shown in Fig.~\ref{fig:6}~\blue{(h)}, the non-Hermiticity still increases the information obtained from measurements and improves the corresponding sensitivity bound.
In simple terms, NHQS based on \textit{Scheme I}, the stronger the non-Hermiticity, the higher the sensitivity, which means the superior QS performance; while NHQS based on \textit{Scheme II} is inferior to HQS.

Consequently, the introduced \A~system for practically constructing the \PT-symmetric system based on \textit{Scheme I} provides additional estimated-parameter information which indeed enhances the coding ability of the parameter generator in the parameterized process.
In contrast, based on \textit{Scheme II}, the total estimated-parameter information remains invariant in the parameterized process, and the dissipation leads the information of the system to flow into the external environment, thereby the coding ability of the parameter generator is counteracted.
This demonstrates that the advantages of NHQS over HQS deeply depend on the extra degrees of freedom introduced to construct NH quantum sensors whether carry additional estimated-parameter information.

In addition, a recent work presented a novel perspective of dissipative QFI based on the Liouvillian parameterized process \cite{Peng2023arXiv}.
They proposed that the LEP provides a more accurate elucidation of the singularity of open systems compared to the HEP, which also explains that the EP does not always improve measurement precision.
However, regardless of the measurement precision that can be obtained near either HEP or LEP, the information obtained from measurements in a dissipative system is generally lower than that without dissipation and lost completely during prolonged evolution.
Although we calculated the conventional QFI of a dissipative system using the NH Hamiltonian of which HEP coincides with LEP, our discussion focuses on NHQS with superior performance over HQS and aims to explore the restriction on the performance improvement of NHQS.

Besides, a theoretical study recently proposed fundamental sensitivity limits for NH quantum sensors \cite{PhysRevLett.131.160801}, we also concur with their viewpoint that when comparing the performance of quantum sensors, it is necessary to fix the quantum resources consumed by these sensors.
Hence, in their study, the considered NH systems do not increase the amount of estimated-parameter information during the parameterized process, while our work centers on investigating whether superior performance can exist in NHQS.
We demonstrated that if the introduced degrees of freedom contain the estimated-parameter information, it will enhance the coding ability of the parameter generator during the parameterized process and make the existence of superiority of NHQS; otherwise, it will weaken the parameter coding ability.
Based on this, we will further consider quantum resources consumed in NHQS.

Of course, we indeed expect the performance of QS to be effectively improved by utilizing the characteristics of NH quantum sensors.
In consequence, our attention would be devoted to \textit{Scheme I} which provides a superior indicator of sensitivity in NHQS.
In the following, from the perspective of quantum resources, our goal is to determine whether there exists the additional requirement of information resources accompanied by the performance improvement of NHQS and ascertain the corresponding resource utilization rate, which has a profound influence on practically taking advantage of NHQS.

(iv) \textit{Information costs and sensitivity losses related to the post-selection.--}
Within the framework of resource theory, based on conceptions of QFI $\mathcal{I}_\omega$ and sensitivity bound $\delta\omega$ in Eqs.~\eqref{I4Lpost}~and~\eqref{CRB_I}, we introduce the following definitions. The information cost $\xi_\mathcal{I_\omega}$ is associated with the post-selection and represents the utilization rate of information resources:
\begin{eqnarray}
\label{cost}
\xi_\mathcal{I_\omega}=\frac{\Delta\mathcal{I}_\omega}{\mathcal{I}_\omega^\mathrm{4d}}=\frac{\mathcal{I}_\omega^\mathrm{4d}-\mathcal{I}_\omega^\mathrm{subs}}{\mathcal{I}_\omega^\mathrm{4d}},
\end{eqnarray}
where $\mathcal{I}_\omega^\mathrm{subs}$ corresponds to the presence of post-selection, while $\mathcal{I}_\omega^\mathrm{4d}$ corresponds to the absence of post-selection.
According to the function of post-selection, the lower the information cost $\xi_\mathcal{I_\omega}$ is, the less redundant information is required, which indicates more available estimated-parameter information encoded in the quantum state.
When $\xi_\mathcal{I_\omega}=1$ represents that the post-selection destroys the coding ability of the parameter generator and depletes all the information resources;
while $\xi_\mathcal{I_\omega}=0$ denotes the absence of redundant information (i.e., the resource utilization rate is $100\%$).
And the corresponding sensitivity losses related to the post-selection is
\begin{eqnarray}
\label{loss}
\zeta_{\delta\omega}=\frac{\delta\omega_\mathrm{4d}}{\delta\omega_\mathrm{subs}}=\sqrt{\frac{\mathcal{I}_\omega^\mathrm{subs}}{\mathcal{I}_\omega^\mathrm{4d}}},
\end{eqnarray}
the ratio of sensitivity bounds $\delta\omega_\mathrm{4d}$ and $\delta\omega_\mathrm{subs}$ directly demonstrates the sensitivity losses caused by the post-selection, a lower $\zeta_{\delta\omega}$ indicates a higher utilization rate of resource consumption, and $\zeta_{\delta\omega}=1$ represents the absence of sensitivity losses, while $\zeta_{\delta\omega}=0$ is the sensitivity total loss (i.e., the resource utilization rate is $0\%$).

\begin{figure*}[!htbp]
\centering
\includegraphics[width=0.98\textwidth]{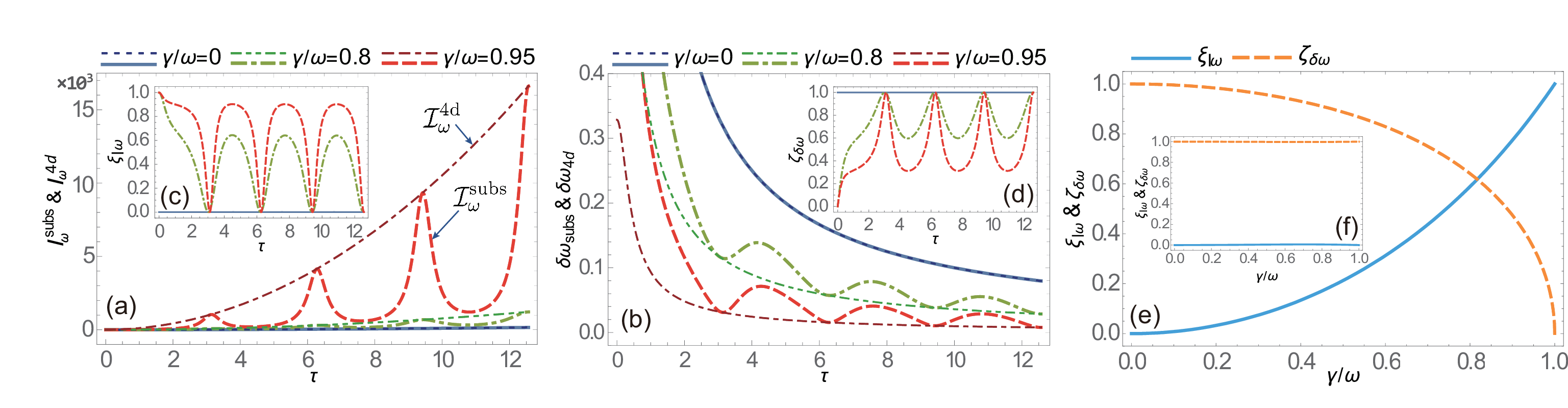}
\caption{(Color online)
QFI $\mathcal{I}_\omega$ in Eq.~\eqref{I4Lpost} and sensitivity bound $\delta\omega$ in Eq.~\eqref{CRB_I} of \textit{Scheme I} with respect to the estimated parameter $\omega$ versus $\tau$ for different $\gamma/\omega$ with the averaged-measure scale $N=1$.
Taking a repeated-averaged measurement:
\textbf{(a)} Comparison of QFI $\mathcal{I}_\omega^\mathrm{subs}$ (thicker-lighter curves) and $\mathcal{I}_\omega^\mathrm{4d}$ (thinner-darker curves);
\textbf{(b)} Comparison of sensitivity bound $\delta\omega_\mathrm{subs}$ (thicker-lighter curves) and $\delta\omega_\mathrm{4d}$ (thinner-darker curves).
Performing the the post-selection:
\textbf{(c)} Information costs $\xi_\mathcal{I_\omega}$ in Eq.~\eqref{cost} versus $\tau$ for different $\gamma/\omega$;
\textbf{(d)} Sensitivity losses $\zeta_{\delta\omega}$ in Eq.~\eqref{loss} versus $\tau$ for different $\gamma/\omega$.
Information costs $\xi_\mathcal{I_\omega}$ and sensitivity losses $\zeta_{\delta\omega}$ with the post-selection versus $\gamma/\omega$ at \textbf{(e)} $\tau=5$ and \textbf{(f)} $\tau=2\pi$.
}
\label{fig:7}
\end{figure*}

Focusing on \textit{Scheme I} which possesses superiority in QS, Fig.~\ref{fig:7}~\blue{(a)} indicates that QFI $\mathcal{I}_\omega^\mathrm{subs}$ (thicker-lighter curves) obtained from the measurements is indeed decreased by the post-selection, compared with $\mathcal{I}_\omega^\mathrm{4d}$ (thinner-darker curves).
Obviously, the stronger the non-Hermiticity, the larger QFI, but the more redundant information is required to be encoded into the parameterized state as shown in Fig.~\ref{fig:7}~\blue{(c)}.
Such an additional requirement of information distinctly increases resource consumption during the parameterized process, such as the extra energy usage \cite{ACSPhoton.9.1554}, which poses an extra challenge for the practical implementation of ultra-precision NHQS and imposes restrictions on its practicable application.
Meanwhile, Fig.~\ref{fig:7}~\blue{(b)} shows that the post-selection increases the costs of information resources and sensitivity losses.
The stronger the non-Hermiticity, the more sensitivity loss shown in Fig.~\ref{fig:7}~\blue{(d)}.
Particularly, at EP ($\gamma/\omega=1$), Fig.~\ref{fig:7}~\blue{(e)} shows that the post-selection depletes all information resources ($\xi_\mathcal{I_\omega}=1$) and completely eliminates sensitivity ($\zeta_\mathcal{I_\omega}=0$).
This demonstrates that, under the post-selection, the resource utilization rate is $0\%$ at EP, which makes the improved performance of NHQS meaningless for practical applications.
In simple terms, although the non-Hermiticity in the \PT-symmetric system indeed makes a superior performance of NHQS over its Hermitian counterpart ($\gamma=0$, blue curves in Figs.~\ref{fig:7}~\blue{(a)~(b)}), it also introduces more information resource requirements under the post-selection.

Consequently, based on \textit{Scheme I}, the practical implementation of performance improvement of NHQS is primarily restricted by the post-selection related to the additional consumption of information resources accompanied by a low utilization rate, even the \PT-symmetric quantum sensor possesses a superior indicator in QS.

Besides, it is easy to determine that the susceptibility $S_\omega=\partial_\omega\rho_\omega^{ii}$ in Eq.~\eqref{S_omega} corresponds to one of the diagonal elements of SLD operator matrix $\partial_\omega\rho_\omega$ in Eq.~\eqref{SLD}.
Meanwhile, SLD operator is a crucial component in the definition of QFI in Eq.~\eqref{F} which quantifies the sensitivity.
The superior performance of NHQS is revealed by the precision of parameter estimation is improved by the non-Hermiticity without involving EP (in Fig.~\ref{fig:6}~\blue{(f)}) and the susceptibility diverges at EP (in Fig.~\ref{fig:5}~\blue{(f)}).
However, it seems counterintuitive that, under the post-selection, the susceptibility diverges at EP, at which sensitivity is eliminated with a resource utilization rate of $0\%$ (in Fig.~\ref{fig:7}~\blue{(e)}).
This contradiction can be attributed to the periodicity in the evolution behavior of \PT-symmetric system.
Particularly, by setting $\tau=2\pi$, $\zeta_{\delta\omega}=1$ also can be obtained even without involving the EP (in Fig.~\ref{fig:7}~\blue{(f)}), which instead indicates that the resource utilization rate is $100\%$.
Note that the main objective here is to determine the potential costs or losses associated with the performance improvement of NHQS, hence we discussed the additional requirement of information resources without considering the special case of (half-) periodic points $\tau=n\pi~(n=1,2,3...)$.

Moreover, there indeed exists a noteworthy feature in EPQS, considering the periodic (half-periodic) points with a vanished success rate (failure rate) at the EP, a unique phenomenon arises: QFI diverges at the EP (i.e., the sensitivity bound approaches to zero) accompanied by the resource utilization rate reaches $100\%$.
Except for the success (or failure) rate of post-selection periodically vanishes and the eigenvectors degenerate at the EP, the mechanism behind this unique phenomenon remains unrevealed.
This is worthwhile to further research on how to realize ultra-precision NHQS in the way of the high efficiency and low consumption of quantum resources.

\section{\label{sec4} Conclusions}
In conclusion, we have investigated the advantages and restrictions of \PT-symmetric QS utilizing two schemes based on the trapped-ion platform.
Our results show that the superior performance of NHQS depends on whether the extra degrees of freedom introduced to practically construct the \PT-symmetric system carry additional information resources.
Moreover, the advantages of NHQS are primarily restricted by the post-selection in practical application, such as the post-selection comes with plenty of potential resource consumption with a low utilization rate, even based upon a scheme with a superior indicator in QS.
This work will provide theoretical guidance for the next stage of experimental efforts to realize the ultra-precision \PT-symmetric QS on the trapped-ion platform.

\begin{acknowledgments}
This work is supported by the National Natural Science Foundation of China (Grant Nos. 11904402, 12074433, 12004430, 12174447, 12174448 and 12365006).
Y.Y.W. is grateful for the help from Professor Mao-Fa Fang.
\end{acknowledgments}

\appendix
\section{\label{apA} More details of \textit{Scheme~I}}

\textit{\PT-symmetric system.--} \PT-symmetric Hamiltonian considered in this paper:
\begin{eqnarray}
\label{A_HPT}
H_\mathcal{PT}=\frac{\omega}{2}\sigma_x+\mathrm{i}\frac{\gamma}{2}\sigma_z=\frac{\omega}{2}
\begin{pmatrix}
\mathrm{i}\gamma/\omega&1\\
1&-\mathrm{i}\gamma/\omega\\
\end{pmatrix},
\end{eqnarray}
the ratio $\gamma/\omega$ is usually regarded as the non-Hermiticity in the system described by $H_\mathcal{PT}$.
The spontaneous \PT-symmetry breaking is generated by the interplay of the tunable gain-loss rate $\gamma$ and the coupling rate $\omega$,
when $\gamma/\omega\in(0,1)$, $H_\mathcal{PT}$ is in the \PT-symmetric phase with real eigenvalues,
while $\gamma/\omega\in(1,\infty)$, $H_{\mathcal{PT}}$ is in the \PT-symmetry broken phase with a pair of complex conjugate eigenvalues.
And $\gamma/\omega=1$ usually represents the EP of $H_\mathcal{PT}$, at which the eigenvalues transition from being real to complex conjugate, accompanied by the degeneracy of the eigenstates.
When $\gamma=0$, represented an absence of non-Hermiticity, the \PT-symmetric system degrades into a (Hermitian) Rabi oscillator with coupling rate $\omega$.
Basing on the eigenvalue equation
\begin{subequations}
\label{A_Eigensystem}
\begin{eqnarray}
\label{A_EigenEq}
H_\mathcal{PT}|\mathrm{E}_\pm\rangle=\mathrm{E}_\pm|\mathrm{E}_\pm\rangle,
\end{eqnarray}
the eigenvalues of $H_\mathcal{PT}$ is given by
\begin{eqnarray}
\label{A_Eigenvalues}
\mathrm{E}_\pm=\pm\kappa/2,
\end{eqnarray}
where $\kappa=\sqrt{\omega^2-\gamma^2}$.
According to the standard Dirac inner product $\langle\mathrm{E}_\pm|\mathrm{E}_\mp\rangle=0$, the orthogonal eigenvectors are given as
\begin{eqnarray}
\label{A_Eigenvectors}
|\mathrm{E}_\pm\rangle
=\frac{1}{\omega}
\begin{pmatrix}
\mathrm{i}\gamma\pm\kappa\\
\omega\\
\end{pmatrix}.
\end{eqnarray}
\end{subequations}
Non-Hermitian Hamiltonians ($H_\mathrm{NH}\neq H_\mathrm{NH}^{\dag}$) always can be decomposed into Hermitian and anti-Hermitian parts as $H_\mathrm{NH}=H_{+}+H_{-}$ with $H_\pm=\pm H_\pm^\dag$.
Decomposing the \PT-symmetric Hamiltonian $H_\mathcal{PT}$ in Eq.~\eqref{A_HPT} into
\begin{eqnarray}
H_+&=&\frac{\omega}{2}\sigma_x;~~H_-=-\mathrm{i}\Gamma=\mathrm{i}\frac{\gamma}{2}\sigma_z,
\end{eqnarray}
and solving the NH dynamics equation \cite{PhysRevLett.109.230405}:
\begin{eqnarray}
\label{A_nHeq}
\dot{\rho}_t=-\mathrm{i}\left[H_+,\rho_t\right]-\left\{\Gamma,\rho_t\right\}+2\Tr\left(\rho_t\Gamma\right)\rho_t.
\end{eqnarray}
However, the NH dynamics Eq.~\eqref{A_nHeq} for a normalized $\rho_t$ is usually hard to solve analytically, its solution also can be expressed as a unitary-like form:
\begin{eqnarray}
\label{A_Rt1}
\rho_t=\frac{U_\mathcal{PT}\rho_0U_\mathcal{PT}^\dag}{\Tr\left(U_\mathcal{PT}\rho_0U_\mathcal{PT}^\dag\right)}
=\begin{pmatrix}
\rho_t^{11}&\rho_t^{12}\\
\rho_t^{21}&\rho_t^{22}\\
\end{pmatrix},
\end{eqnarray}
where $U_\mathcal{PT}=\exp(-\mathrm{i}H_\mathcal{PT}t)$ is the non-unitary time-evolution operator of \PT-symmetric system and $\rho_0$ is an arbitrary initial state.
Considering an initial pure state $\rho_0=|\psi_0\rangle\langle\psi_0|$:
\begin{eqnarray}
|\psi_0\rangle
=\begin{pmatrix}
a_0\\
b_0\\
\end{pmatrix},
\end{eqnarray}
the solution in Eq.~\eqref{A_Rt1} can be reduced to
\begin{subequations}
\label{A_Psi}
\begin{eqnarray}
\label{A_Psit}
\nonumber|\psi_t\rangle&=&\frac{U_\mathcal{PT}|\psi_0\rangle}{\sqrt{\langle\psi_0|U_\mathcal{PT}^\dag U_\mathcal{PT}|\psi_0\rangle}}\\
&=&\mathrm{c_n}\cdot U_\mathcal{PT}|\psi_0\rangle
=\begin{pmatrix}
a_t\\
b_t\\
\end{pmatrix},
\end{eqnarray}
where
\begin{eqnarray}
\label{A_Psiab}
\nonumber a_t&=&\mathrm{c_n}\left[a_0\cos\left(\frac{\kappa t}{2}\right)+\frac{1}{\kappa}\left(a_0\gamma-\mathrm{i}b_0\omega\right)\sin\left(\frac{\kappa t}{2}\right)\right];~~~~~~~\\
b_t&=&\mathrm{c_n}\left[b_0\cos\left(\frac{\kappa t}{2}\right)-\frac{1}{\kappa}\left(b_0\gamma+\mathrm{i}a_0\omega\right)\sin\left(\frac{\kappa t}{2}\right)\right],
\end{eqnarray}
and the renormalized factor $\mathrm{c_n}$ is given by
\begin{eqnarray}
\label{A_Psic}
\nonumber\frac{1}{\mathrm{c_n}^2}&=&\Tr\left(U_\mathcal{PT}|\psi_0\rangle\langle\psi_0|U_\mathcal{PT}^\dag\right)=\langle\psi_0|U_\mathcal{PT}^\dag U_\mathcal{PT}|\psi_0\rangle\\
\nonumber&=&\left[b_0\cos\left(\frac{\kappa t}{2}\right)-\frac{1}{\kappa}\left(\mathrm{i}a_0\omega+b_0\gamma\right)\sin\left(\frac{\kappa t}{2}\right)\right]\\
\nonumber&\times&\left[b_0^*\cos\left(\frac{\kappa t}{2}\right)+\frac{1}{\kappa}\left(\mathrm{i}a_0^*\omega-b_0^*\gamma\right)\sin\left(\frac{\kappa t}{2}\right)\right]\\
\nonumber&+&\left[a_0\cos\left(\frac{\kappa t}{2}\right)+\frac{1}{\kappa}\left(a_0\gamma-\mathrm{i}b_0\omega\right)\sin\left(\frac{\kappa t}{2}\right)\right]~~~~~~\\
&\times&\left[a_0^*\cos\left(\frac{\kappa t}{2}\right)+\frac{1}{\kappa}\left(a_0^*\gamma+\mathrm{i}b_0^*\omega\right)\sin\left(\frac{\kappa t}{2}\right)\right].~~~~~~~
\end{eqnarray}
\end{subequations}
The corresponding density operator $\rho_t$ in Eq.~\eqref{A_Rt1} can be represented as
\begin{subequations}
\begin{eqnarray}
\rho_t&=&|\psi_t\rangle\langle\psi_t|
=\begin{pmatrix}
a_t\\
b_t\\
\end{pmatrix}
\begin{pmatrix}
a_t^*~b_t^*
\end{pmatrix},
\end{eqnarray}
and the matrix elements of normalized $\rho_t$ are
\begin{eqnarray}
\label{A_elements}
\nonumber\rho_t^{11}&=&|a_t|^2;~~\rho_t^{22}=1-\rho_t^{11}=|b_t|^2;\\
\rho_t^{12}&=&a_tb_t^*;~~\rho_t^{21}=\left(\rho_t^{12}\right)^*=b_ta_t^*.
\end{eqnarray}
\end{subequations}
The initial state in this paper is chosen as
\begin{eqnarray}
\label{A_Psi0+}
|\psi_0\rangle&=&|+\rangle_y
=\frac{\sqrt{2}}{2}\begin{pmatrix}
1\\
\mathrm{i}\\
\end{pmatrix},
\end{eqnarray}
i.e., $a_0=1/\sqrt{2};~b_0=\mathrm{i}/\sqrt{2}$, where $|\pm\rangle_y=(|1\rangle\pm\mathrm{i}|2\rangle)/\sqrt{2}$ are the eigenvectors of the Pauli matrix $\sigma_y$.\\

\textit{Scheme~I: Naimark-dilated quantum system.}-- Quantum simulating the \PT-symmetric Hamiltonian through introducing an auxiliary system.
Utilizing the Naimark-dilation theory, embedding the \PT-symmetric subspace into a larger Hermitian space with an auxiliary subspace, the \PT-symmetric non-unitary dynamics can be mapped from a unitary dynamics in the enlarged system, the process is decomposed into the following two steps.\\

\textbf{Step1}:~\emph{Utilizing the Naimark-dilation theory.}

Constructing a Hermitian metric operator $\eta=\eta^\dag$ based on the pseudo-Hermiticity condition \cite{JMathPhys.43.2814}:
\begin{eqnarray}
\eta H_\mathcal{PT}=\left(\eta H_\mathcal{PT}\right)^\dag=H_\mathcal{PT}^\dag\eta^\dag=H_\mathcal{PT}^\dag\eta.
\end{eqnarray}
With the basis $\{|i\rangle\}$ ($i=1,2,3,4$), the \PT-symmetric subsystem (\PT-sub) $|\psi_t\rangle$ is denoted with $\{|1\rangle,|2\rangle\}$, and the auxiliary subsystem (\A-sub) $|\chi_t\rangle$ is denoted with $\{|3\rangle,|4\rangle\}$.
According to the \PT~inner product \cite{PhysRevLett.89.270401}:
\begin{eqnarray}
\left(\mu,\nu\right)=\left(\mathcal{PT}\mu\right)\cdot\nu,
\end{eqnarray}
where $\mathcal{P}=\sigma_x$ is the parity operator and $\mathcal{T}$ is the complex conjugate operator, the complete (orthonormalized) eigenvectors $|E_\pm\rangle$ of $H_\mathcal{PT}$ are satisfied with $(E_\pm,E_\mp)=0$ and
\begin{subequations}
\label{A_inner}
\begin{eqnarray}
\label{A_PTinner}
\left(E_\pm,E_\pm\right)=\pm1,
\end{eqnarray}
while turning back to $|\mathrm{E}_\pm\rangle$ accorded to the standard Dirac inner product given by Eq.~\eqref{A_Eigenvectors}:
\begin{eqnarray}
\label{A_Diracinner}
\left(\mathrm{E}_\pm,\mathrm{E}_\pm\right)=\left(\mathcal{PT}|\mathrm{E}_\pm\rangle\right)\cdot|\mathrm{E}_\pm\rangle=\pm\frac{2\kappa}{\omega},
\end{eqnarray}
\end{subequations}
and setting $f$ as the normalized factor of transformation:
\begin{eqnarray}
&&\frac{1}{f^2}=\frac{2\kappa}{\omega}\Longrightarrow f=\frac{1}{\sqrt{2\kappa/\omega}}.
\end{eqnarray}
Arranging $|E_\pm\rangle=f|\mathrm{E}_\pm\rangle$ into the row matrix:
\begin{eqnarray}
\nonumber\Phi&=&\left(|E_+\rangle~|E_-\rangle\right)=f\left(|\mathrm{E}_+\rangle~|\mathrm{E}_-\rangle\right)\\
&=&\frac{1}{\sqrt{2\kappa\omega}}
\begin{pmatrix}
\mathrm{i}\gamma+\kappa&\mathrm{i}\gamma-\kappa\\
\omega&\omega\\
\end{pmatrix},
\end{eqnarray}
and the metric operator is defined by
\begin{eqnarray}
\nonumber\eta&=&(\Phi\Phi^\dagger)^{-1}=\frac{\omega}{\kappa}\left(\mathbb{I}+\frac{\gamma}{\omega}\sigma_y\right)\\
&=&\frac{\omega}{\kappa}
\begin{pmatrix}
1&-\mathrm{i}\gamma/\omega\\
\mathrm{i}\gamma/\omega&1\\
\end{pmatrix}.
\end{eqnarray}
Regarding the metric operator $\eta$ as a synchronization link of dynamics between \A-sub (non-normalized $|\chi_t\rangle$) and \PT-sub (normalized $|\psi_t\rangle$ in Eq.~\eqref{A_Psit}):
\begin{subequations}
\begin{eqnarray}
|\chi_t\rangle=\eta|\psi_t\rangle
=\begin{pmatrix}
c_t\\
d_t\\
\end{pmatrix},
\end{eqnarray}
where
\begin{eqnarray}
\nonumber c_t&=&\frac{1}{\kappa}\left(\omega a_t-\mathrm{i}\gamma b_t\right);\\
d_t&=&\frac{1}{\kappa}\left(\mathrm{i}\gamma a_t+\omega b_t\right).
\end{eqnarray}
\end{subequations}

There are three serial perspectives for understanding the dynamics of enlarged Hermitian system: (a) constructing a total wave function of the enlarged Hermitian system by directly connecting the dynamics of \PT-sub and \A-sub by the synchronization link $\eta$ as Eq.~\eqref{A_Psi}; (b) figuring out the totally unitary time-evolution operator of the enlarged Hermitian system and obtaining its purely unitary evolution processing as Eqs.~\eqref{A_Psi2}~and~\eqref{A_Rho}; (c) expanding a Hermitian Hamiltonian to govern the dynamics of enlarged Hermitian system as Eq.~\eqref{A_H4d}.

Embedding the \PT-symmetric space as a NH subspace $\mathcal{H_{PT}}$ into a larger Hermitian space $\mathcal{H}_\mathrm{H(4d)}=\mathcal{H_{PT}}_{(\mathrm{2d})}\oplus\mathcal{H}_{\mathcal{A}(\mathrm{2d})}$ with an auxiliary subspace $\mathcal{H_A}$ \cite{PhysRevLett.101.230404}, and the wave function of enlarged Hermitian system is constructed as
\begin{subequations}
\label{A_Psi}
\begin{eqnarray}
\nonumber|\Psi_t\rangle
&=&\mathrm{C_n}\left(
\begin{bmatrix}
1\\
0\\
\end{bmatrix}
\otimes|\psi_t\rangle+
\begin{bmatrix}
0\\
1\\
\end{bmatrix}
\otimes|\chi_t\rangle\right)\\
&=&\mathrm{C_n}
\begin{pmatrix}
\psi_t\\
\chi_t\\
\end{pmatrix}
=\left(\tilde{a}_t~\tilde{b}_t~\tilde{c}_t~\tilde{d}_t\right)^\mathrm{T},
\end{eqnarray}
with the renormalized factor
\begin{eqnarray}
\mathrm{C_n}=\frac{1}{\sqrt{\langle\psi_t|\psi_t\rangle+\langle\chi_t|\chi_t\rangle}}
=\frac{1}{\sqrt{1+|c_t|^2+|d_t|^2}},~~~~~~~~~
\end{eqnarray}
elements of $|\Psi_t\rangle$ with $|\psi_0\rangle=|+\rangle_y$ are given as
\begin{eqnarray}
\nonumber\tilde{a}_t&=&\mathrm{c_{4d(+)}}\left[\kappa\cos\left(\kappa t/2\right)+\left(\omega+\gamma\right)\sin\left(\kappa t/2\right)\right];\\
\nonumber\tilde{b}_t&=&\mathrm{ic_{4d(-)}}\left[\left(\omega-\gamma\right)\cos\left(\kappa t/2\right)-\kappa\sin\left(\kappa t/2\right)\right];\\
\nonumber\tilde{c}_t&=&\mathrm{c_{4d(+)}}\left[\left(\omega+\gamma\right)\cos\left(\kappa t/2\right)+\kappa\sin\left(\kappa t/2\right)\right];\\
\tilde{d}_t&=&\mathrm{ic_{4d(-)}}\left[\kappa\cos\left(\kappa t/2\right)-\left(\omega-\gamma\right)\sin\left(\kappa t/2\right)\right].~~~~~~
\end{eqnarray}
\end{subequations}
where $\mathrm{c_{4d(\pm)}}=[4\omega(\omega\pm\gamma)]^{-1/2}$.

The unitary time evolution of enlarged Hermitian system according to Eq.~\eqref{A_Psi} also can be expressed as
\begin{subequations}
\label{A_Psi2}
\begin{eqnarray}
\nonumber|\Psi_t\rangle
&=&\mathrm{C_n}
\begin{pmatrix}
\psi_t\\
\chi_t\\
\end{pmatrix}
=\mathrm{C_n}
\begin{pmatrix}
\psi_t\\
\eta\psi_t\\
\end{pmatrix}
=\mathrm{C_n}
\begin{pmatrix}
\mathrm{c_n}\cdot U_{\mathcal{PT}}\psi_0\\
\mathrm{c_n}\cdot\eta U_{\mathcal{PT}}\eta^{-1}\chi_0\\
\end{pmatrix}\\
\nonumber&=&\mathrm{C_n}\mathrm{c_n}
\begin{pmatrix}
U_{\mathcal{PT}}&0\\
0&\eta U_{\mathcal{PT}}\eta^{-1}\\
\end{pmatrix}
\begin{pmatrix}
\psi_0\\
\chi_0\\
\end{pmatrix}\\
\nonumber&=&\mathrm{\tilde{C}_n}
\begin{pmatrix}
U_{\mathcal{PT}}&0\\
0&U_{\mathcal{A}}\\
\end{pmatrix}
\begin{pmatrix}
\psi_0\\
\chi_0\\
\end{pmatrix}
=\mathrm{\tilde{C}_n}\cdot U_\mathrm{H}
\begin{pmatrix}
\psi_0\\
\chi_0\\
\end{pmatrix}\\
&=&U_\mathrm{4d}|\Psi_0\rangle,
\end{eqnarray}
where $U_\mathcal{A}=\eta U_\mathcal{PT}\eta^{-1}$ with $U_\mathcal{PT}=\exp(-\mathrm{i}H_\mathcal{PT}t)$, and the purely unitary time-evolution operator of enlarged Hermitian system is
\begin{eqnarray}
U_\mathrm{4d}=(\mathrm{C_n}\mathrm{c_n})\cdot U_\mathrm{H}=\mathrm{\tilde{C}_n}\cdot U_\mathrm{H},
\end{eqnarray}
with the renormalized factor
\begin{eqnarray}
\mathrm{\tilde{C}_n}=1/\sqrt{\langle\psi_0|U_\mathcal{PT}^\dag U_\mathcal{PT}|\psi_0\rangle+\langle\chi_0|U_\mathcal{A}^\dag U_\mathcal{A}|\chi_0\rangle},~~~~~~~
\end{eqnarray}
and the unitary-like $U_\mathrm{H}$ is directly obtained by applying the synchronization link of subsystems dynamics $\eta$:
\begin{eqnarray}
\label{A_UH}
U_\mathrm{H}
=\begin{pmatrix}
U_{\mathcal{PT}}&0\\
0&U_{\mathcal{A}}\\
\end{pmatrix}
=\begin{pmatrix}
U_\mathcal{PT}&0\\
0&\eta U_\mathcal{PT}\eta^{-1}\\
\end{pmatrix}.
\end{eqnarray}
\end{subequations}
With the Naimark-dilation theory, we have deduced the unitary time-evolution operator $U_\mathrm{4d}U_\mathrm{4d}^\dag=\mathbb{I}$ related with the purely Hermitian Hamiltonian $H_\mathrm{4d}=H_\mathrm{4d}^\dag$ which is obtained from $U_\mathrm{4d}=\exp(-\mathrm{i}H_\mathrm{4d}t)$.
And the purely unitary evolution of enlarged Hermitian system (denoted by the normalized density operator $\tilde{\rho}_\mathrm{4d}$) is represented as
\begin{subequations}
\label{A_Rho}
\begin{eqnarray}
\tilde{\rho}_\mathrm{4d}&=&|\Psi_t\rangle\langle\Psi_t|=U_{\mathrm{4d}}|\Psi_0\rangle\langle\Psi_0| U_{\mathrm{4d}}^\dag=U_{\mathrm{4d}}\tilde{\rho}_{0}U_{\mathrm{4d}}^\dag,~~~~~~~
\end{eqnarray}
where $\tilde{\rho}_0=|\Psi_0\rangle\langle\Psi_0|$ is the initial state related to $|\Psi_0\rangle=(\psi_0~\chi_0)^\mathrm{T}$, here $|\chi_0\rangle=\eta|\psi_0\rangle$ is the initial state of \A-sub and $|\psi_0\rangle$ for \PT-sub given by Eq.~\eqref{A_Psi0+}, and the matrix diagonal elements (i.e., population of $|i\rangle$ ($i=1,2,3,4$)) of normalized $\tilde{\rho}_\mathrm{4d}$ are
\begin{eqnarray}
\label{A_elements4d}
\nonumber\tilde{\rho}_\mathrm{4d}^{11}&=&|\tilde{a}_t|^2=\left[\omega-\gamma\cos\left(\kappa t\right)+\kappa\sin\left(\kappa t\right)\right]/4\omega;\\
\nonumber\tilde{\rho}_\mathrm{4d}^{22}&=&|\tilde{b}_t|^2=\left[\omega-\gamma\cos\left(\kappa t\right)-\kappa\sin\left(\kappa t\right)\right]/4\omega;\\
\nonumber\tilde{\rho}_\mathrm{4d}^{33}&=&|\tilde{c}_t|^2=\left[\omega+\gamma\cos\left(\kappa t\right)+\kappa\sin\left(\kappa t\right)\right]/4\omega;\\
\tilde{\rho}_\mathrm{4d}^{44}&=&|\tilde{d}_t|^2=\left[\omega+\gamma\cos\left(\kappa t\right)-\kappa\sin\left(\kappa t\right)\right]/4\omega,~~~~~~
\end{eqnarray}
and other matrix off-diagonal elements:
\begin{eqnarray}
\nonumber\tilde{\rho}_\mathrm{4d}^{12}&=&\left(\tilde{\rho}_\mathrm{4d}^{21}\right)^*=\mathrm{i}\left[\gamma-\omega\cos\left(\kappa t\right)\right]/4\omega;\\
\nonumber\tilde{\rho}_\mathrm{4d}^{13}&=&\left(\tilde{\rho}_\mathrm{4d}^{31}\right)^*=\left[\kappa+\omega\sin\left(\kappa t\right)\right]/4\omega;\\
\nonumber\tilde{\rho}_\mathrm{4d}^{14}&=&\left(\tilde{\rho}_\mathrm{4d}^{41}\right)^*=-\mathrm{i}\left[\kappa\cos\left(\kappa t\right)+\gamma\sin\left(\kappa t\right)\right]/4\omega;\\
\nonumber\tilde{\rho}_\mathrm{4d}^{23}&=&\left(\tilde{\rho}_\mathrm{4d}^{32}\right)^*=\mathrm{i}\left[\kappa\cos\left(\kappa t\right)-\gamma\sin\left(\kappa t\right)\right]/4\omega;\\
\nonumber\tilde{\rho}_\mathrm{4d}^{24}&=&\left(\tilde{\rho}_\mathrm{4d}^{42}\right)^*=\left[\kappa-\omega\sin\left(\kappa t\right)\right]/4\omega;\\
\tilde{\rho}_\mathrm{4d}^{34}&=&\left(\tilde{\rho}_\mathrm{4d}^{43}\right)^*=-\mathrm{i}\left[\gamma+\omega\cos\left(\kappa t\right)\right]/4\omega.~~~~~~
\end{eqnarray}
\end{subequations}

Note that, the unitary time evolution of the enlarged Hermitian system in this paper is directly characterised by connecting the dynamics of two subsystems by the synchronization link $\eta$, and we have verified this time-evolution process is consistent with that of Ref.~\cite{PhysRevLett.101.230404} which proposed a unitary evolution governed by an expanded dimension Hamiltonian, the corresponding $H_\mathrm{4d}$ based on the \PT-symmetric system in this paper is
\begin{eqnarray}
\label{A_H4d}
\nonumber H_\mathrm{4d}=&&f^2\left[\mathbb{I}\otimes\left(H_\mathcal{PT}\eta^{-1}+\eta H_\mathcal{PT}\right)\right]\\
&&+f^2\left[\mathrm{i}\sigma_y\otimes\left(H_\mathcal{PT}-H_\mathcal{PT}^\dag\right)\right].
\end{eqnarray}

\textbf{Step2}:~\emph{Executing the post-selection.}

Considering the post-selection performed on the enlarged Hermitian system, the success and failure rate of post-selection are respectively denoted as
\begin{eqnarray}
\nonumber p_\mathrm{suc}&=&P_1+P_2;\\
p_\mathrm{fail}&=&1-p_\mathrm{suc}=P_3+P_4,
\end{eqnarray}
where $P_{i(t)}=\tilde{\rho}_\mathrm{4d}^{ii}$ ($i=1,2,3,4$) is the population of enlarged Hermitian system, $\tilde{\rho}_\mathrm{4d}^{ii}$ are the matrix diagonal elements of $\tilde{\rho}_\mathrm{4d}$ in Eq.~\eqref{A_elements4d}.
The renormalized population for \PT-sub:
\begin{subequations}
\label{A_P}
\begin{eqnarray}
\label{A_P1P2}
\tilde{P}_1&=&\frac{P_1}{P_1+P_2}=\rho_\mathcal{PT}^{11};~~\tilde{P}_2=\frac{P_2}{P_1+P_2}=\rho_\mathcal{PT}^{22},~~~~~
\end{eqnarray}
the corresponding renormalized population for \A-sub:
\begin{eqnarray}
\label{A_P3P4}
\tilde{P}_3&=&\frac{P_3}{P_3+P_4}=\rho_\mathcal{A}^{11};~~~\tilde{P}_4=\frac{P_4}{P_3+P_4}=\rho_\mathcal{A}^{22}.~~~~~~~
\end{eqnarray}
\end{subequations}
Executing the post-selection on the enlarged system $\tilde{\rho}_\mathrm{4d}$, if successful, $\tilde{\rho}_\mathrm{4d}$ is degraded into the \PT-symmetric system $\rho_\mathcal{PT}$:
\begin{subequations}
\label{A_4L}
\begin{align}
\rho_\mathcal{PT}=\rho_t
=\frac{|\psi_t\rangle\langle\psi_t|}{\Tr\left(|\psi_t\rangle\langle\psi_t|\right)}
=\begin{pmatrix}
\rho_t^{11}&\rho_t^{12}\\
\rho_t^{21}&\rho_t^{22}\\
\end{pmatrix},
\end{align}
here $\rho_t^{ij}~(i,j=1,2)$ of $\rho_t$ in Eq.~\eqref{A_Rt1} with $|\psi_0\rangle=|+\rangle_y$:
\begin{eqnarray}
\nonumber\rho_\mathcal{PT}^{11}&=&\rho_t^{11}=\left[1+\mathrm{c_{pt}}\kappa\sin\left(\kappa t\right)\right]/2;\\
\nonumber\rho_\mathcal{PT}^{12}&=&\rho_t^{12}=\mathrm{i}\mathrm{c_{pt}}\left[\gamma-\omega\cos\left(\kappa t\right)\right]/2;\\
\nonumber\rho_\mathcal{PT}^{21}&=&\rho_t^{21}=\left(\rho_t^{12}\right)^*=-\mathrm{i}\mathrm{c_{pt}}\left[\gamma-\omega\cos\left(\kappa t\right)\right]/2;\\
\rho_\mathcal{PT}^{22}&=&\rho_t^{22}=1-\rho_t^{11}=\left[1-\mathrm{c_{pt}}\kappa\sin\left(\kappa t\right)\right]/2,~~~~~~
\end{eqnarray}
\end{subequations}
where $\mathrm{c_{pt}}=[\omega-\gamma\cos(\kappa t)]^{-1}$; if failure, $\tilde{\rho}_\mathrm{4d}$ is degraded into the \A~system $\rho_\mathcal{A}$:
\begin{subequations}
\label{A_4L2}
\begin{eqnarray}
\rho_\mathcal{A}=\frac{|\chi_t\rangle\langle\chi_t|}{\Tr\left(|\chi_t\rangle\langle\chi_t|\right)}
=\begin{pmatrix}
\rho_t^{22}&\rho_t^{21}\\
\rho_t^{12}&\rho_t^{11}\\
\end{pmatrix},
\end{eqnarray}
here $\rho_t^{ij}~(i,j=1,2)$ of $\rho_t$ in Eq.~\eqref{A_Rt1} with $|\psi_0\rangle=|-\rangle_y$:
\begin{eqnarray}
\nonumber\rho_\mathcal{A}^{11}&=&\rho_t^{22}=\left[1+\mathrm{c_{a}}\kappa\sin\left(\kappa t\right)\right]/2;\\
\nonumber\rho_\mathcal{A}^{12}&=&\rho_t^{21}=-\mathrm{i}\mathrm{c_{a}}\left[\gamma+\omega\cos\left(\kappa t\right)\right]/2;\\
\nonumber\rho_\mathcal{A}^{21}&=&\rho_t^{12}=\left(\rho_t^{21}\right)^*=\mathrm{i}\mathrm{c_{a}}\left[\gamma+\omega\cos\left(\kappa t\right)\right]/2;\\
\rho_\mathcal{A}^{22}&=&\rho_t^{11}=1-\rho_t^{22}=\left[1-\mathrm{c_{a}}\kappa\sin\left(\kappa t\right)\right]/2,~~~~~~
\end{eqnarray}
\end{subequations}
where $\mathrm{c_{a}}=[\omega+\gamma\cos(\kappa t)]^{-1}$.
Eqs.~\eqref{A_4L}~and~\eqref{A_4L2} indicate that, because of the full synchronization link $\eta$ with the \PT-sub, the \A-sub indeed possesses the \PT-symmetry.
In consequence, two subsystems possess an anti-mirror-symmetric correlation (see Figs.~\ref{fig:4d}~\blue{(e) (f)}), and the enlarged Hermitian system combined by two fully synchronized \PT-symmetric subsystems can be regarded as a pseudo-dual-\PT-symmetric system, and EPs of \PT-sub and \A-sub overlap at $\gamma/\omega=1$.

\section{\label{apB} More details of \textit{Scheme~II}}

\textit{Scheme~II: Effective non-Hermitian Hamiltonian of open quantum system.--} Quantum simulating the \PT-symmetric Hamiltonian by introducing an external environment.
As we know, the dynamics of open quantum systems obey the Lindblad master equation:
\begin{subequations}
\begin{eqnarray}
\dot{\rho}_t=\mathcal{L}\rho_t,
\end{eqnarray}
with the dynamics generator $\mathcal{L}$ which is an arbitrary, linear or non-linear, Liouvillian super-operator:
\begin{align}
\mathcal{L}(\cdot)\equiv-\mathrm{i}\left[H_0,(\cdot)\right]+\mathcal{D}(\cdot),
\end{align}
and the dissipative term $\mathcal{D}(\cdot)$ can be expressed as
\begin{eqnarray}
\mathcal{D}(\cdot)=J(\cdot)J^\dag-\frac{1}{2}\left\{J^\dag J,(\cdot)\right\},
\end{eqnarray}
\end{subequations}
where $J$ is the quantum jump operator.

\textbf{Method~1}: \textit{Effective rule and re-normalization.}

In order to study the \PT-symmetric system with $H_\mathcal{PT}$, considering a dissipative two-level system described by an NH Hamiltonian with the basis $\{|i\rangle\}~(i=1,2)$:
\begin{eqnarray}
\label{B_Heff}
\nonumber H_\mathrm{eff}&=&\frac{\omega}{2}\sigma_x-\mathrm{i}\gamma~|2\rangle\langle2|\\
\nonumber&=&H_\mathcal{PT}-\mathrm{i}\frac{\gamma}{2}\mathbb{I}\\
&=&\frac{\omega}{2}
\begin{pmatrix}
0&1\\
1&-\mathrm{i}2\gamma/\omega\\
\end{pmatrix},
\end{eqnarray}
where $\omega$ is the coupling rate, $\gamma$ is the tunable decay rate.

Firstly, such a dissipative two-level system also can be regarded as an effective description of a three-level system, with the basis $\{|i\rangle\}~(i=1,2,3)$, the coherent transition is denoted by $|1\rangle\leftrightarrow|2\rangle$ with the coupling rate $\omega$, the dissipation of two-level system is represented by $|2\rangle\rightarrow|3\rangle$ with the tunable decay rate $\gamma$, and no transition between $|1\rangle\nleftrightarrow|3\rangle$.
The dynamics of the three-level system can be described by the Lindblad master equation with a Liouvillian super-operator $\mathcal{L}$:
\begin{eqnarray}
\label{B_Leq}
\nonumber\dot{\varrho}_t&=&\mathcal{L}\varrho_t=-\mathrm{i}\left[H_0,\varrho_t\right]+\mathcal{D}\varrho_t\\
&=&-\mathrm{i}\left[\frac{\omega}{2}\sigma_x,\varrho_t\right]+\left(J\varrho_tJ^\dag-\frac{1}{2}\left\{J^\dag J,\varrho_t\right\}\right),~~~~
\end{eqnarray}
where $\varrho_t$ is the normalized density operator of three-level system, $H_0=\omega\sigma_x/2$ is the coherent transition Hamiltonian, and $J=\sqrt{\gamma}~|3\rangle\langle2|$ is the jump operator.
Through solving the Lindblad master Eq.~\eqref{B_Leq} with the initial state $|\tilde{\psi}_0\rangle=|\tilde{+}\rangle_y=(|1\rangle+\mathrm{i}|2\rangle+0|3\rangle)/\sqrt{2}$ and $\tilde{\varrho}_0=|\tilde{\psi}_0\rangle\langle\tilde{\psi}_0|$, we can obtain the density operator $\varrho_t$ and its matrix elements:
\begin{eqnarray}
\label{B_elements}
\nonumber\varrho_t^{11}&=&\mathrm{c_{3L}}\left[\omega-\gamma\cos\left(\kappa t\right)+\kappa\sin\left(\kappa t\right)\right];\\
\nonumber\varrho_t^{12}&=&\mathrm{ic_{3L}}\left[\gamma-\omega\cos\left(\kappa t\right)\right];\\
\nonumber\varrho_t^{21}&=&\left(\varrho_t^{12}\right)^*=-\mathrm{ic_{3L}}\left[\gamma-\omega\cos\left(\kappa t\right)\right];\\
\nonumber\varrho_t^{22}&=&\mathrm{c_{3L}}\left[\omega-\gamma\cos\left(\kappa t\right)-\kappa\sin\left(\kappa t\right)\right];\\
\nonumber\varrho_t^{33}&=&1-(\varrho_t^{11}+\varrho_t^{22})=1-2\mathrm{c_{3L}}\left[\omega-\gamma\cos\left(\kappa t\right)\right];\\
\varrho_t^{13}&=&\varrho_t^{31}=\varrho_t^{23}=\varrho_t^{32}=0,
\end{eqnarray}
where $\mathrm{c_{3L}}=\mathrm{e}^{-\gamma t}[2(\omega-\gamma)]^{-1}$.

Secondly, according to the effective rule of open quantum systems, the dynamics generated by $H_\mathrm{eff}$ in Eq.~\eqref{B_Heff} can be depicted through reducing dimension of Eq.~\eqref{B_Leq}: modifying the quantum jump term $|2\rangle\rightarrow|3\rangle$ with the loss rate $\gamma$ in $\mathcal{D}\varrho_t$ as the decay term $-\mathrm{i}\gamma|2\rangle\langle2|$ in $H_\mathrm{eff}$, and then the lower-dimension Lindblad master equation without considering quantum jumps can be used to describe the dynamics of  dissipative two-level system:
\begin{subequations}
\label{B_effdanamics}
\begin{eqnarray}
\label{B_Lreff}
\dot{\varrho}_\mathrm{eff}=\mathcal{L}\varrho_\mathrm{eff}=-\mathrm{i}\left[H_\mathrm{eff},\varrho_\mathrm{eff}\right],
\end{eqnarray}
and the corresponding differential equations system:
\begin{eqnarray}
\label{B_elementsdot}
\nonumber\dot{\varrho}_\mathrm{eff}^{11}&=&\mathrm{i}\frac{\omega}{2}\left(\varrho_\mathrm{eff}^{12}-\varrho_\mathrm{eff}^{21}\right);\\
\nonumber\dot{\varrho}_\mathrm{eff}^{12}&=&\mathrm{i}\frac{\omega}{2}\left(\varrho_\mathrm{eff}^{11}-\varrho_\mathrm{eff}^{22}\right)-\gamma\varrho_\mathrm{eff}^{12};\\
\nonumber\dot{\varrho}_\mathrm{eff}^{21}&=&\mathrm{i}\frac{\omega}{2}\left(\varrho_\mathrm{eff}^{22}-\varrho_\mathrm{eff}^{11}\right)-\gamma\varrho_\mathrm{eff}^{21};\\
\dot{\varrho}_\mathrm{eff}^{22}&=&\mathrm{i}\frac{\omega}{2}\left(\varrho_\mathrm{eff}^{21}-\varrho_\mathrm{eff}^{12}\right)-2\gamma\varrho_\mathrm{eff}^{22}.
\end{eqnarray}
\end{subequations}
Deducing the matrix of Liouvillian super-operator as
\begin{subequations}
\begin{eqnarray}
\mathcal{L}=\begin{pmatrix}
0&\mathrm{i}\omega/2&-\mathrm{i}\omega/2&0\\
\mathrm{i}\omega/2&-\gamma&0&-\mathrm{i}\omega/2\\
-\mathrm{i}\omega/2&0&-\gamma&\mathrm{i}\omega/2\\
0&-\mathrm{i}\omega/2&\mathrm{i}\omega/2&-2\gamma\\
\end{pmatrix},
\end{eqnarray}
and the eigenvalues $\mathrm{E}_{\mathcal{L}(i)}$ ($i=1,2,3,4$) of $\mathcal{L}$ matrix are
\begin{align}
\left\{\mathrm{E}_{\mathcal{L}(i)}\right\}=\left\{-\gamma\pm\mathrm{i}\kappa_{(1)(4)},-\gamma_{(2)(3)}\right\},
\end{align}
\end{subequations}
with the LEP $\gamma/\omega=1$ coincided with the HEP of $H_\mathcal{PT}$.
The solution of Eq.~\eqref{B_effdanamics} equals to matrix elements of the lower-dimension $\varrho_t$ in Eq.~\eqref{B_elements}:
\begin{eqnarray}
\label{B_elements2}
\varrho_\mathrm{eff}^{ij}=\varrho_t^{ij}~(i,j=1,2).
\end{eqnarray}
The solution also can be simply expressed as
\begin{eqnarray}
\label{B_reff}
\varrho_\mathrm{eff}=U_\mathrm{eff}\varrho_0U_\mathrm{eff}^\dag,
\end{eqnarray}
where $U_\mathrm{eff}=\exp(-\mathrm{i}H_\mathrm{eff}t)$.
In order to simulate the characteristic of energy balance of gain and loss in the \PT-symmetric system, by directly adding a gain term $\mathrm{i}\gamma|1\rangle\langle1|$ on $H_\mathrm{eff}$ of the dissipative two-level system whose energy is only with loss and without gain, the artificial \PT-symmetric system can be obtained:
\begin{eqnarray}
\label{B_rPT}
\varrho_\mathcal{PT}=\mathrm{e}^{\gamma t}\varrho_\text{eff}.
\end{eqnarray}

Finally, performing the re-normalization on $\varrho_\mathcal{PT}$:
\begin{align}
\label{B_Reff1}
\rho_\mathcal{PT}=\frac{\varrho_\mathcal{PT}}{\Tr(\varrho_\mathcal{PT})}=\frac{\mathrm{e}^{\gamma t}\varrho_\mathrm{eff}}{\Tr(\mathrm{e}^{\gamma t}\varrho_\mathrm{eff})}=\frac{\varrho_\mathrm{eff}}{\Tr(\varrho_\mathrm{eff})}=\rho_\mathrm{eff},
\end{align}
and then we realize the construction of \PT-symmetric system $\rho_\mathcal{PT}$ governed by $H_\mathcal{PT}$ with the dissipative two-level system $\rho_\mathrm{eff}$ dominated by an effective NH Hamiltonian $H_\mathrm{eff}$.\\

\textbf{Method~2}: \textit{Post-selection.}

From the perspective of the post-selection, after solving the Lindblad master Eq.~\eqref{B_Leq} and obtaining the density matrix $\varrho_t$ of three-level system, we also can achieve \PT-symmetric system $\rho_\mathcal{PT}$ through directly performing the post-selection on the three-level system $\varrho_t$:
\begin{align}
\label{B_Reff2}
\rho_\mathcal{PT}=\frac{1}{\varrho_t^{11}+\varrho_t^{22}}
\begin{pmatrix}
\varrho_t^{11}&\varrho_t^{12}\\
\varrho_t^{21}&\varrho_t^{22}\\
\end{pmatrix}
=\begin{pmatrix}
\rho_t^{11}&\rho_t^{12}\\
\rho_t^{21}&\rho_t^{22}\\
\end{pmatrix},
\end{align}\\
where $\rho_t^{ij}~(i,j=1,2)$ of $\rho_t$ in Eq.~\eqref{A_Rt1} with $|\psi_0\rangle=|+\rangle_y$, and the success rate of post-selection $p_\mathrm{suc}=\varrho_t^{11}+\varrho_t^{22}$ actually denotes the norm of $\varrho_\mathrm{eff}$ in Eq.~\eqref{B_elements2}.

\bibliography{QS_ref_2}

\end{document}